\documentclass[11pt,a4paper]{article}
\usepackage[utf8]{inputenc}
\usepackage{amsmath}
\usepackage{amsthm}
\usepackage{amsfonts}
\usepackage{amssymb}
\usepackage{graphicx}

\usepackage{multirow}
\usepackage{natbib}
\usepackage{graphicx}
\usepackage[ruled]{algorithm2e}
\usepackage{xcolor}
\newtheorem{lemma}{Lemma}
\newtheorem{proposition}{Proposition}
\newtheorem{remark}{Remark}
\newtheorem{corollary}{Corollary}
\usepackage{hyperref}

\begin{document}

\title{\textbf{Choice of trimming proportion and number of clusters in robust clustering based on trimming}
\thanks{This research is partially supported by the Spanish Ministerio de Ciencia e Innovaci\'{o}n, grant PID2021-128314NB-I00. The work also benefits from the High Performance Computing (HPC) facility of the University of Parma.}}

\author{\textit{Luis Angel Garc\'{\i}a-Escudero} \\
    Department of Statistics and Operational Research and
    IMUVA\\ University of Valladolid\\
    \textit{Christian Hennig} \\
    Department of Statistical Sciences ``Paolo Fortunati''
    \\ University of Bologna\\
    \textit{Agust\'{\i}n Mayo-Iscar} \\
    Department of Statistics and Operational Research and
    IMUVA\\ University of Valladolid\\
    \textit{Gianluca Morelli} \\
    Department of Economics and Management and RoSA\\
    University of Parma\\
    \textit{Marco Riani} \\
    Department of Economics and Management and RoSA\\
     University of Parma}

\date{}
\maketitle

\begin{abstract}
So-called ``classification trimmed likelihood curves'' have been proposed as a useful heuristic tool to determine the number of clusters and trimming proportion in trimming-based robust clustering methods. However, these curves needs a careful visual inspection, and this way of choosing parameters requires subjective decisions. This work is intended to provide theoretical background for the understanding of these curves and the elements involved in their derivation. Moreover, a parametric bootstrap approach is presented in order to automatize the choice of parameter more by providing a reduced list of ``sensible'' choices for the parameters. The user can then pick a solution that fits their aims from that reduced list.

\end{abstract}

\section{Introduction}\label{s1}
Outliers are known to be potentially detrimental for many statistical procedures and, consequently, the use of more robust alternatives is often advisable. This is also the case in cluster analysis, for which some robust clustering approaches have been developed. These proposals are designed to better resist a certain fraction of outlying observations. Moreover, clustered outliers are known to be one of the most challenging types of contamination (even for robust statistical procedures) and, thus, applying an unified treatment for both outliers and clusters should make perfect sense. Some references on robust clustering are \cite{GarG10,Rit14,63_GarGHen16}.

In this work we will focus on trimming-based approaches for robust clustering, firstly introduced in \cite{CueG97} with the trimmed $k$-means. There, a so-called ``impartial'' trimming approach was proposed. The term impartial means that the data set itself will tell us what fraction $\alpha$ of observations should be trimmed. This idea also underlies high breakdown-point robust methods \citep{RosL87} such as, for instance, the MCD (Minimum Covariance Determinant) estimator. Throughout this manuscript, we focus on TCLUST \citep{GarG08} as our reference robust clustering method. TCLUST may be seen as an extension of both the trimmed $k$-means and the MCD. It allows for impartial trimming and the detection of elliptical clusters with flexible shapes, whereas trimmed $k$-means looks for spherical clusters with the same scatter only. TCLUST is implemented in the R-package \texttt{tclust} \citep{FriG12} and the \texttt{FSDA} Matlab toolbox \citep{24c}.

When using TCLUST, the number of clusters $k$ is required, and there is also the need of choosing the proportion $\alpha$ of trimmed observations.
The choice of a suitable number of clusters $k$ is one of the most relevant, and also complex, issues in clustering. It is known to be a hard problem since in practice it depends on the purpose of the clustering; clusterings at finer or coarser granularity may be required. Moreover, finding the ``true'' number of clusters in probability mixture model-based clustering (for example identifying the number of clusters with the number of Gaussian mixture components) is not a perfectly well-posed problem, as model assumptions are never perfectly fulfilled, and any distribution that does not follow the model precisely can be approximated better adding more mixture components, see the discussion in \cite{Hennig15handbook}. Furthermore, in the robust clustering problem where some observations can be outliers, there is ambiguity between small clusters and groups of outliers, making the problem of choosing the number of clusters $k$ even more difficult, see \cite{Hennig2023}. This means that user input (such as a decision on how small a homogeneous group of points is seen as a group of outliers rather than a cluster, which may also depend on the homogeneity, i.e., within-group variability) is required in order to distinguish between several potentially existing legitimate clustering solutions on the same data. The final decision will typically depend on the meaning of the data and the aim of clustering.

On the other hand, subjective user decisions are unpopular and problematic in science. Existing information is not always sufficient for motivating such decisions, and the ideal of scientific objectivity implies that conclusions should be reproducible, and should not depend on the individual researcher. This ideal cannot be achieved with the ill-defined problem of estimating the number of clusters in potential presence of outliers, see \cite{Hennig2023} for more discussion. What we propose here is an automatic method that equips the user with a number of ``sensible'', well justified, and formally well defined solutions that still allow for different possible trade-offs between clusters and trimmed outliers, so that the user still has to make the final decision in case a single clustering is required.

There is ambiguity between a classification of observations as outliers or as clusters that are either very small or have very large within-cluster variation. As already commented, the choices of $k$ and $\alpha$ are clearly interrelated, and in order to deal with them, it makes sense to take into account the scatter of the clusters. TCLUST allow the user to determine the maximum allowed scatter difference between clusters through an ``eigenvalue ratio'' constraint. This involves an additional parameter $c$, which will interact with the choices of $\alpha$ and $k$.

Given a sample $\mathcal{X}_n=\{x_1,...,x_n\}$ with $x_i\in\mathbb{R}^p$ 
(which we later assume will be the result of an independent identically distributed (i.i.d.) sample of size $n$ from an unknown distribution $P$), $k\geq 1$, $\alpha \in [0,1)$ and $c\geq 1$, the TCLUST method is defined through the maximization of the trimmed classification likelihood
\begin{equation}\label{e_0}
\sum_{j=1}^k \sum_{i\in R_j}
    \log (\pi_j \phi(x_i;\mu_j,\Sigma_j)),
\end{equation}
where $\phi(\cdot;\mu,\Sigma)$ stands for the probability density function of the $p$-variate normal distribution (with mean parameter $\mu$ and scatter matrix $\Sigma$), $\{R_0,R_1,\allowbreak ...,R_k\}$ is a partition of the indexes $\{1,2,...,n\}$ such that $\#R_0=[n \alpha]$ ($R_0$ indicates the set of trimmed observations). Moreover, the maximization of (\ref{e_0}) is subject to the eigenvalue ratio constraint
\begin{equation}\label{e_1}
M_n/m_n\leq c,
\end{equation}
where $M_n=\max_{j=1,...,k}\max_{l=1,...,p} \lambda_l(\Sigma_{j})$ and $m_n=\min_{j=1,...,k}\min_{l=1,...,p}\allowbreak \lambda_l(\Sigma_{j})$
are, respectively, the largest and the smallest of the eigenvalues of the scatter matrices $\Sigma_j$ $j=1,...,k$. The $\pi_j\geq 0$ have to fulfill $\sum_{j=1}^k \pi_j=1$. Throughout this work, to simplify the notation, we will suppress the dependence on the constant $c$ in the presentation unless strictly necessary, even though $c$ in (\ref{e_1}) plays a key role. When $c$ is close to 1, the situation is close to looking for spherical clusters with the same scatter. Moreover, $c<\infty$ is key for achieving a mathematically well-defined problem, because it will
avoid scatter matrix degeneracy (i.e., components with $|\Sigma_j|$ close to 0), that makes the objective function (\ref{e_0}) unbounded, and the detection of meaningless ``spurious clusters'' that can happen if a small set of observations randomly produces a very small covariance matrix eigenvalue.

Analyzing the evolution of the maximum value attained by (\ref{e_0}) when moving $k$ and $\alpha$ in a controlled way has been found informative for deriving sensible values for $k$ and $\alpha$. In fact, this was the approach already suggested in \cite{GarG03} through the use of the ``$k$-variograms'' for trimmed $k$-means. Regarding TCLUST, monitoring of the ``classification trimmed likelihood'' curves (\texttt{ctlcurves}) was proposed in \cite{GarG11}. Let $\mathcal{L}^{\Pi}(\alpha,k;\mathcal{X}_n)$ denote the maximum value attained in the constrained maximization of  (\ref{e_0}) for a given fixed constant $c$. The \texttt{ctlcurves} method is based on the careful analysis of the curves
\begin{equation}\label{ctlcurves}
(k,\alpha) \mapsto \mathcal{L}^{\Pi}(\alpha,k;\mathcal{X}_n)
\end{equation}
for $k=1,...,K_{\max}$ and $\alpha=1,...,\alpha_{\max}$. The heuristic argument given in \cite{GarG11} is that a large $t_{k,\alpha}^n=\mathcal{L}^{\Pi}(\alpha,k+1;\mathcal{X}_n) - \mathcal{L}^{\Pi}(\alpha,k;\mathcal{X}_n)$ indicates a clear benefit increasing the number of clusters from $k$ to $k+1$ for the specific trimming level $\alpha$. This heuristic provides a useful exploratory tool for determining sensible values for $k$ and $\alpha$. However, unfortunately, the determination of those sensible values requires a careful visual examination of the curves and forces the user to make subjective decisions about whether $t_{k,\alpha}^n$ can be considered large enough or not.

\begin{figure}\label{f1}
\centering
\hspace*{0.5cm}{\scriptsize \textsf{(a)}}

\includegraphics[clip,width=8cm, height=6.25cm]{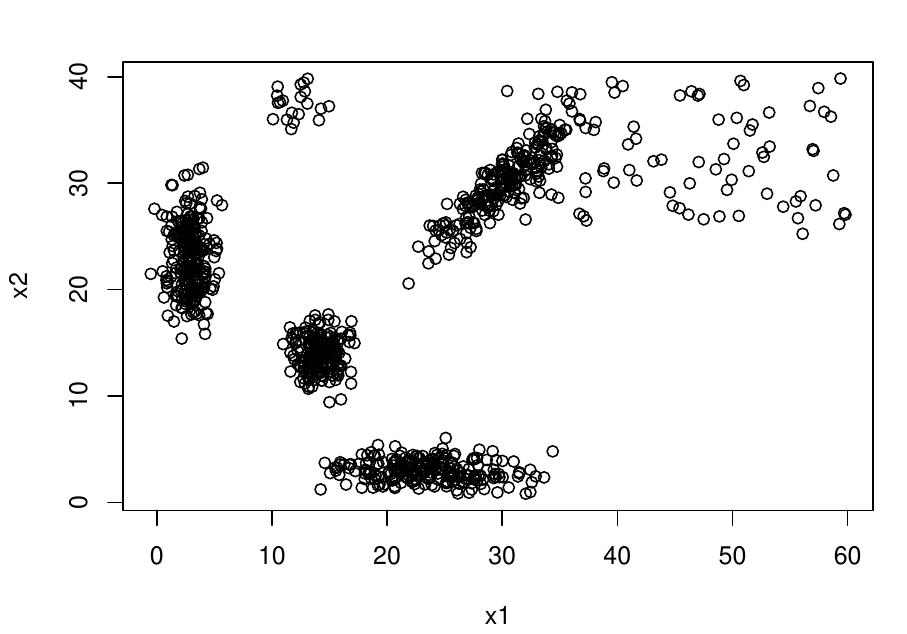}

\hspace*{0.5cm}{\scriptsize \textsf{(b)}}

\includegraphics[clip,width=7.85cm, height=6.25cm]{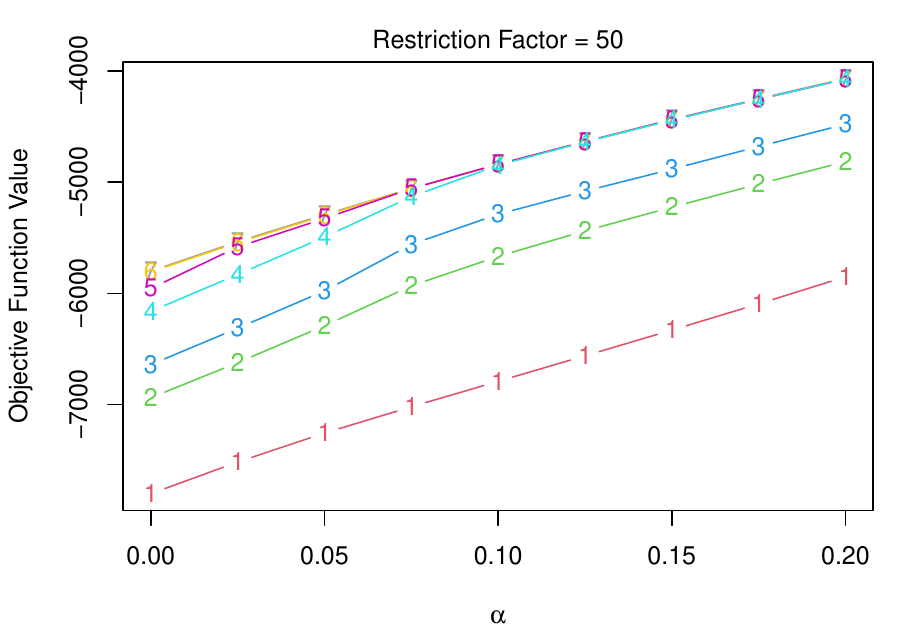}
\caption{(a) Simulated data set, (b) associated \texttt{ctlcurves}.}
\end{figure}

Figure \ref{f1}(a) shows a simulated data set and its corresponding \texttt{ctlcurves} in (b) for $c=50$, $K_{\max}=7$ and $\alpha_{\max}=0.2$. A well-trained user may interpret these curves as suggesting the three solutions that will be shown in Figure \ref{f2}. This choice is however not straightforward because of the sample variability of these curves that can be observed when zooming into specific parts of Figure \ref{f1}(b). There could also be inaccuracies due to issues with convergence and local optima of the applied algorithm, which is based on random initialization and concentration steps \citep{FriG13}. This type of algorithmic instability, although acknowledged, will not be taken into account in this work, but it deserves attention.

One of the main purposes of this work is to provide more theoretical support for the use of the \texttt{ctlcurves}, which were only introduced as a heuristic in \cite{GarG11}. Section \ref{s2} is devoted to the better understanding of the elements involved in the \texttt{ctlcurves} approach and its justification. 
Section \ref{s3} presents a new parametric bootstrap approach that can be used to automatize the interpretation of the \texttt{ctlcurves} taking into account sample variability. The parametric bootstrap approach provides the user with a list of sensible solutions for further investigation. Section \ref{s4} shows some simulated illustrative examples for this approach. A comparative simulation study is presented in Section \ref{s5}. Its application to two real data examples is shown in Section \ref{s6}. Section \ref{s7} concludes the paper.

\section{Theoretical results}\label{s2}

Given $\mathcal{X}_n=\{x_1,...,x_n\}\subset \mathbb{R}^p$, $k\in \mathbb{N}$, $\alpha\in[0,1)$ and fixed $c\geq 1$, define
\begin{equation}\label{e1}
\mathcal{L}^{\Pi}(\alpha,k;\mathcal{X}_n)= \max_{\substack{R=\{R_j\}_{j=0}^k \in
    \mathcal{R}_{\alpha}\\ \theta=\{\theta_j \}_{j=1}^k \in \Theta_c ,
   \pi=(\pi_1,...,\pi_k) \in \Pi_k}} \sum_{j=1}^k \sum_{i\in R_j}
    \log (\pi_j \phi(x_i;\theta_j)),
\end{equation}
where $\mathcal{R}_{\alpha}$ is the set of possible partitions $R= \{R_j \}_{j=0}^k$ with $\cup_{j=0}^k R_j = \{1,...,n\}$, $R_r \cap
R_s=\emptyset$ for $r\neq s$ and $\#R_0=[n \alpha]$, and
$$\Theta_c=\{ \theta=\{\theta_j \}_{j=1}^k:\theta_j=(\mu_j,\Sigma_j)\text{ such that
}\Sigma_j\text{'s satisfy } M_n/m_n\leq c\},$$
where  $M_n$ and $m_n$ are, respectively, the largest and the smallest  eigenvalues of the scatter matrices $\Sigma_{j}$, and the component weights $\pi=(\pi_1,...,\pi_k)$ are in the set
$\Pi_k=\big\{\pi\in [0,1]^k: \sum_{j=1}^k \pi_j=1\big\}.$
Existence of these optimal parameters and the optimal partition were proven in \cite{GarG08} under mild assumptions. The notation $\mathcal{L}^{\Pi}(\alpha,k;\mathcal{X}_n)$ in (\ref{e1}) omits its dependence on the eigenvalue ratio constraint $c$.

For fixed $k$ and $\alpha$ define
$$
t_{k,\alpha}^n= \mathcal{L}^{\Pi}(\alpha,k+1;\mathcal{X}_n) - \mathcal{L}^{\Pi}(\alpha,k;\mathcal{X}_n).
$$
It is clear that $t_{k,\alpha} \geq 0$ because equality can be reached by choosing one $R_j$ as empty set. Additionally, it was heuristically argued in \cite{GarG11} that $t_{k,\alpha}^n$ tends to be small when $k$ is a ``sensible'' choice for the given trimming level $\alpha$, which justifies the interest of monitoring the \texttt{ctlcurves} in (\ref{ctlcurves}) as exploratory tool for determining suitable choices of the parameters $k$ and  $\alpha$.

We now introduce the population version of the classification trimmed likelihood curves (for a more detailed presentation of these concepts see \cite{GarG08} and \cite{GarG11}) depending on an unknown underlying probability measure $P$ that we assume to have generated the observed sample $\mathcal{X}_n$. Given $P$, define
 \begin{equation*}
    \mathcal{L}_{\alpha,k}^{\Pi}(P)= \max_{\substack{Z=\{Z_j\}_{j=0}^k \in
    \mathcal{Z}_{\alpha}\\ \theta=\{\theta_j \}_{j=1}^k \in \Theta_c ,
   \pi=(\pi_1,...,\pi_k) \in \Pi_k}} P \bigg[ \sum_{j=1}^k I_{Z_j}(\cdot)
    \log (\pi_j \phi(\cdot;\theta_j)) \bigg],
\end{equation*}
where $\mathcal{Z}_{\alpha}$ denotes the set of all possible partitions
of $\mathbb{R}^p$, i.e., $\cup_{j=0}^k Z_j\equiv \mathbb{R}^p$, $Z_j \cap
Z_{j'}=\emptyset$ for $j\neq j'$ such that
$
P\big[\cup_{j=1}^k Z_j\big]= 1 - \alpha,
$
and $P[f(\cdot)]=\int f(x) dP(x)$.

The partition $Z=\{Z_j\}_{j=0}^k \in \mathcal{Z}_{\alpha}$ maximizing
\begin{equation}\label{tclust_tar}
P \bigg[ \sum_{j=1}^k I_{Z_j}(\cdot)
    \log (\pi_j \phi(\cdot;\theta_j)) \bigg]
\end{equation}
for fixed parameters $\theta\in \Theta_c$, and $\pi\in \Pi$
can be obtained by considering
$$Z_0=Z_0(\theta,\pi;P)=\mathbb{R}^p \setminus B(\theta,\pi;P)$$
and
$$
Z_j=Z_j(\theta,\pi;P)=Z_j(\theta,\pi)\cap B(\theta,\pi;P), \text{ for }j=1,...,k,$$
where
$$
Z_j(\theta,\pi)=\{x: \pi_j\phi(x;\theta_j)\geq \pi_r\phi(x;\theta_r)\text{ for }r=1,...,k\},
$$
$$
B(\theta,\pi;P)=\bigg\{ x\in \mathbb{R}^p: \max_{j=1,...,k} \pi_j\phi(x;\theta_j)\geq R(\theta,\pi;P)\bigg\},
$$
and $R(\theta,\pi;P)$ is the $\alpha$-quantile of $\max_{j=1,...,k} \pi_j\phi(X;\theta_j)$ for a random variable $X$ with distribution $P$.
To simplify the presentation and avoid introducing trimming functions \citep{GarG08}, we consider  absolutely continuous distributions $P$ so that sets $B(\theta,\pi;P)$ with $P[B(\theta,\pi;P)]=1-\alpha$ do always exist for any $\theta$ and $\pi$.
We again omit the dependence on $c$ in the notation $\mathcal{L}_{\alpha,k}^{\Pi}(P)$. We also have that $ \mathcal{L}_{\alpha,k+1}^{\Pi}(P) \geq \mathcal{L}_{\alpha,k}^{\Pi}(P)$, because we can always set $Z_j=\emptyset$ for one or more values of $j$.

Given $\mathcal{X}_n=\{x_1,...,x_n\}$, if $P_n$ denotes the associated empirical measure $P_n=\frac{1}{n}\sum_{i=1}^n \delta_{x_i}$, with $\delta_x$ the Dirac measure concentrated on $x$, then
$$\mathcal{L}^{\Pi}(\alpha,k;\mathcal{X}_n)=\mathcal{L}_{\alpha,k}^{\Pi}(P_n),$$  and consequently,
$
t_{k,\alpha}^n= \mathcal{L}_{\alpha,k+1}^{\Pi}(P_n) - \mathcal{L}_{\alpha,k}^{\Pi}(P_n).
$

The following remark is a direct consequence of the consistency result proven as Proposition 1 in \cite{GarG11}:

\begin{remark}\label{propo0} If $\{P_n\}_{n=1}^{\infty}$ is the sequence of empirical distributions associated to i.i.d. random observations from $P$, then
$$
\mathcal{L}_{\alpha,k+1}^{\Pi}(P_n) - \mathcal{L}_{\alpha,k}^{\Pi}(P_n)  \rightarrow \mathcal{L}_{\alpha,k+1}^{\Pi}(P) - \mathcal{L}_{\alpha,k}^{\Pi}(P), \text{ almost surely for }n\rightarrow \infty
$$
for any $k\in \mathbb{N}$ and $\alpha\in[0,1)$ and fixed $c$.

The consistency result only requires very mild assumptions on $P$. For instance, no moment conditions are needed, unless in the $\alpha=0$ case where $P|\cdot|^2 < \infty$ is needed. We only need to avoid pathological cases where the distribution $P$ is not concentrated on a finite number of points after trimming a certain fraction $\alpha$ of the probability mass. These moment conditions and lack of concentration on a finite number of points will trivially hold for the distribution $P=P_0$ that will be considered later.
\end{remark}

Remark \ref{propo0} implies that $t_{k,\alpha}^n$ will approximate 0, whenever $\mathcal{L}_{\alpha,k}^{\Pi}(P)=\mathcal{L}_{\alpha,k+1}^{\Pi}(P)$. Given that the interpretation of the \texttt{ctlcurves} depends on $t_{k,\alpha}^n$, this consistency result motivates the relevance of the theoretical study of changes in  $\mathcal{L}_{\alpha,k}^{\Pi}(P)$, when moving $k$, for some distributions $P=P_0$ that could be of interest in the clustering framework.

\begin{remark} It is convenient to introduce weights in $\Pi_k$ for choosing $k$ and $\alpha$. Suppose we use
 \begin{equation*}
    \mathcal{L}_{\alpha,k}(P)= \underset{\{Z_j\}_{j=0}^k \in
    \mathcal{Z}_{\alpha}, \{\theta_j\}_{j=1}^k \in \Theta_c,
    \{\pi_j\}_{j=1}^k \in [0,1]^k}{\max} P \bigg[ \sum_{j=1}^k I_{Z_j}(\cdot)
    \log ( \phi(\cdot;\theta_j)) \bigg],
\end{equation*}
removing the $\pi_j$ weights in (\ref{e1}). Then we can define an analogous statistic $r_{k,\alpha}^n=\mathcal{L}_{\alpha,k+1}(P) - \mathcal{L}_{\alpha,k}(P)
$. However, $r_{k,\alpha}^n$ can never converge to 0 because Proposition 2 in \cite{GarG11} implies that always $\mathcal{L}_{\alpha,k+1}(P) - \mathcal{L}_{\alpha,k}(P)>0$. In this case it is not feasible to define a criterion based on evaluating whether $r_{k,\alpha}^n$ is close to 0 or not. This issue is well known and enforces modified criteria when monitoring changes in the within-clusters sum of squares objective function in $k$-means clustering (which amounts to $c=1$ and $\pi_1=...=\pi_k$) such as, for instance, Calinski-Harabasz \citep{calinski1974dendrite}, or the gap statistic \citep{sugar2003finding}.
\end{remark}

In order to establish some results about $\mathcal{L}_{\alpha,k}^{\Pi}(P_0)$, the following version of the well-known Gibbs' inequality will be useful:

\begin{lemma}\label{le_gibbs} If $P_0$ is an absolutely continuous distribution with probability density functions $f_0$, then
\begin{equation}\label{gibbs}
P_0[\log f_0(\cdot)]\geq P_0[\log f(\cdot)]
\end{equation}
for any other probability density function $f$. \emph{(\ref{gibbs})} is an equality if and only if $f_0=f$ almost $P_0$-everywhere.
\end{lemma}

We will start considering the (easier) untrimmed $\alpha=0$ case and assuming normality for the $k_0$ components defining the underlying distribution $P=P_0$. If $k_0=1$ and $P_0$ only includes one component, we assume that $P_0$ is a $p$-variate normal with parameters $\theta_1^0=(\mu_0,\Sigma_0)$. However,  introducing convenient $P_0$ distributions that could serve as a theoretical framework when assuming $k_0>1$ normally distributed ``components'' is not completely straightforward. We assume that $\pi_j^0>0$ and that $\theta_j^0\neq \theta_{j'}^0$ for every $j,j'\in\{1,...,k_0\}$ where $\theta_j^0=(\mu_j^0,\Sigma_j^0)$ are the parameters defining each of these $k_0$ normally distributed components. Moreover, concerned with clustering problems, in order to be relevant, $P_0$ should have an interpretation in terms of clusters and not necessarily as a mixture of normal components. In view of (\ref{e1}), it makes sense to consider $P_0$ as admitting a probability density function related to $\sum_{j=1}^{k_0} I_{Z_j(\theta^0,\pi^0)}(\cdot) \pi_j^0 \phi(\cdot;\theta_j^0)$. But these functions do not integrate to 1, and it is necessary to normalize them to construct a proper density for $P_0$, which we denote as $f_0$, given by
\begin{equation}\label{e2}
f_0(x)=\eta(\theta^0,\pi^0)\sum_{j=1}^{k_0} I_{Z_j(\theta^0,\pi^0)}(x) \pi_j^0 \phi(x;\theta_j^0)\text{ for }x\in \mathbb{R}^p,
\end{equation}
with a normalizing constant $\eta(\theta^0,\pi^0)$ obtained through the function
$$
\eta(\theta,\pi)=\frac{1}{\sum_{j=1}^k \pi_j p_j\big(Z_j(\theta,\pi);\theta\big)},
$$
with
\begin{equation}\label{e4}
p_j(Z;\theta)=\int_Z\phi(x;\theta_j) dx,\ j=1,...,k,
\end{equation}
for $\theta=\{\theta_j\}_{j=1}^k$. Observe that $p_j(Z;\theta)$ is the probability that a $p$-variate normal with parameters $\theta_j=(\mu_j,\Sigma_j)$, extracted from $\theta$, gives to $Z$.

We trivially have $\eta(\theta,\pi)\ge 1$ for any $\theta$ and $\pi$. Note also that $\eta(\theta^0,\pi^0)$ can be seen as some kind of index of overlap in $P_0$ because it measures to what extent the normal components defined by the $\theta^0$ and $\pi^0$ are more or less overlapped. If the assumed underlying normal components are well-separated, then all $p_j(Z_j(\theta^0,\pi^0);\theta^0)\approx 1$ and thus, consequently, $\eta(\theta^0,\pi^0)\approx 1$. The larger $\eta(\theta^0,\pi^0)$, the more overlapped these $k$ normal components are.

Table \ref{t1} shows the $\eta(\theta^0,\pi^0)$ values measuring overlap for different values of $\theta^0$ and $\pi^0$ for two different normal components in the univariate case to help understanding the effect of the overlap.

 \begin{table}[hb]
 \centering
 \begin{tabular}{lc}
 \hline
 Parameters &  $\eta(\theta^0,\pi^0)$\\
 \hline
 $\pi=(0.5,0.5)$, $\theta_1=(-0.2,1)$ and $\theta_2=(0.2,1)$ & 1.727 \\
 $\pi=(0.5,0.5)$, $\theta_1=(-0.5,1)$ and $\theta_2=(0.5,1)$ & 1.447 \\
 $\pi=(0.1,0.9)$, $\theta_1=(-0.2,0.1)$ and $\theta_2=(0,1)$ & 1.111 \\
 $\pi=(0.02,0.98)$, $\theta_1=(-0.2,0.1)$ and $\theta_2=(0,1)$ & 1.020 \\
 $\pi=(0.5,0.5)$, $\theta_1=(-5,1)$ and $\theta_2=(5,1)$ & 1.000 \\
 $\pi=(0.02,0.98)$, $\theta_1=(-3.8,0.1)$ and $\theta_2=(0,1)$ & 1.000 \\
 \hline
 \end{tabular}
 \caption{Indexes of overlap for different parameter values for $k_0=2$ normal components in $p_0$ when dimension $p=1$.}\label{t1}
 \end{table}

We are now in the situation to establish the following result, which is a direct consequence of the Gibbs' inequality in Lemma \ref{le_gibbs}.

\begin{proposition}\label{proposition1} For $P_0$ with density $f_0$ as in \emph{(\ref{e2})}, we have
\begin{eqnarray}
 &&P_0\bigg[ \sum_{j=1}^{k_0} I_{Z_j(\theta^0,\pi^0)}(\cdot)
    \log (\pi_j^0 \phi(\cdot;\theta_j^0)) \bigg] - P_0 \bigg[ \sum_{j=1}^k I_{Z_j(\theta,\pi)}(\cdot)
    \log (\pi_j \phi(\cdot;\theta_j)) \bigg] \nonumber\\
    &&\geq \log \frac{\eta(\theta,\pi)}{\eta(\theta^0,\pi^0)}   \label{pr1_0}
\end{eqnarray}
for any $\theta$ and $\pi$, and that \emph{(\ref{pr1_0})} is an equality only when $k=k_0$, $\theta=\theta^0$ and $\pi=\pi^0$ or when $k>k_0$ but $\theta_j=\theta_j^0$, $\pi_j=\pi_j^0$, for $j=1,...,k_0$, and $\pi_{k_0+1}=...=\pi_k=0$ (after suitable relabeling).
 \end{proposition}
\noindent \textit{Proof:} The proof follows easily by applying Lemma \ref{le_gibbs} to $P_0$, $P_0$ with density $f_0$, and the density $f$ with
$$
f(x)=\eta(\theta,\pi)\sum_{j=1}^{k} I_{Z_j(\theta,\pi)}(x) \pi_j \phi(x;\theta_j)\text{ for }x\in \mathbb{R}^p,
$$
and rearranging the constant terms. \hspace*{0.2cm}$\square$\medskip

A first consequence of Proposition \ref{proposition1} is the simplest case $k_0=1$ when $P_0$ is a (non-degenerated) $p$-variate normal:

\begin{corollary}\label{col1} If $P_0$ is a $p$-variate normal with mean $\mu_0$ and covariance matrix $\Sigma_0$ (with $|\Sigma_0|\neq 0$), then
$
\mathcal{L}_{0,1}^{\Pi}(P_0)=\mathcal{L}_{0,k}^{\Pi}(P_0)\text{ for any }k\in \mathbb{N}.
$
\end{corollary}
\noindent \textit{Proof:} Recall $\mathcal{L}_{0,1}^{\Pi}(P_0)\leq\mathcal{L}_{0,k}^{\Pi}(P_0)$. It is obvious that, for this $P_0$ and $k=1$, we have $\pi_1=1$, $Z_0\equiv \emptyset$ and $Z_1\equiv \mathbb{R}^p$ so that $
\mathcal{L}_{0,1}^{\Pi}(P_0)=P_0[\log\phi(\cdot;\mu_0,\Sigma_0)]$. Given that $\eta(\theta^0,\pi^0)=1$ and $\eta(\theta,\pi)\geq 1$, the inequality (\ref{pr1_0}) implies
\begin{equation}\label{k_1}
P_0[\log\phi(\cdot;\mu_0,\Sigma_0)] \geq P_0 \bigg[ \sum_{j=1}^k I_{Z_j(\theta,\pi)}(\cdot)
    \log (\pi_j \phi(\cdot;\theta_j)) \bigg],
\end{equation}
and, thus, $\mathcal{L}_{0,1}^{\Pi}(P_0)\geq\mathcal{L}_{0,k}^{\Pi}(P_0)$ for that $P_0$.
\hspace*{0.2cm}$\square$
 \medskip

Let us continue assuming a distribution $P_0$ with the probability density function $f_0$ in (\ref{e2}), but now with $k_0>1$ components. We will follow the same reasoning as in the proof of Corollary \ref{col1}, making use of (\ref{pr1_0}) again. In the ideal case that $\eta(\theta^0,\pi^0)=1$ (i.e., the overlap between components is null), we would have an analogous result to Corollary \ref{col1} stating that $$\mathcal{L}_{0,k}^{\Pi}(P) < \mathcal{L}_{0,k_0}^{\Pi}(P)\text{ when } k<k_0,$$ and that $$\mathcal{L}_{0,k}^{\Pi}(P) = \mathcal{L}_{0,k_0}^{\Pi}(P)\text{ when } k\geq k_0,$$ because, in that case,
$\log\big(\eta(\theta,\pi)/\eta(\theta^0,\pi^0)\big)\geq 0$ since $\eta(\theta,\pi)\geq 1$. However, upon computing (\ref{e2}), it becomes evident that normally distributed components have positive density all over $x\in\mathbb{R}^p$ and, and as a result, we never achieve  $\eta(\theta^0,\pi^0)=1$ exactly. However, the previous argument shows that the behavior would be very close to that in $\mathcal{L}_{0,k}(P_0)$ whenever $\eta(\theta^0,\pi^0)\approx 1$ or, in other words, when $k_0$ normally distributed components in $P_0$ are well separated, with negligible overlaps among them.

Obviously, the true $P_0$ will not always consist of $k_0$ well separated normal components on which the above arguments are elaborated. In those cases, the inequality (\ref{pr1_0}) helps us to understand that when maximizing the population version of TCLUST in expression (\ref{tclust_tar}) under $P=P_0$, considering parameters $\theta$ and $\pi$ associated with very large $\eta(\theta,\pi)$ overlap values is going to be penalized, and it could be better to reduce the number of clusters $k$ detected by leaving empty clusters (i.e., $\pi_j=0$). This penalization of overlap makes sense in cluster analysis, because two strongly overlapped normal components with densities $\phi(\cdot;\mu_j,\Sigma_j)$ and $\phi(\cdot;\mu_{j'}\Sigma_{j'})$ can and in some applications should be considered as generating one single cluster. Many likelihood-ratio test procedures available in the literature (see, e.g., \cite{McLP00,mclachlan2014number}) are based on mixture likelihoods, and thus prefer fitting two different normal components as long as $\mu_j\neq \mu_{j'}$  or $\Sigma_j\neq \Sigma_{j'}$, even though the associated components are  heavily overlapped. In clustering this is often not desirable. Therefore an approach based on classification trimmed likelihoods with weights, implying the penalization discussed above, is better suited to such clustering problems than approaches based on mixture likelihoods or on classification likelihoods without weights. This more clustering-oriented flavor of our proposal also is behind the use of the well-known ICL criterion in  \cite{BieC00} instead of the BIC.

If we were able to incorporate the $\eta(\theta,\pi)$ in the objective function for every possible choice of $\theta$ and $\pi$, this would define a methodology that would not penalize overlap. It could potentially asymptotically select the correct $\theta^0$ and $\pi^0$ when sampling from a distribution $P_0$ with a density as given in (\ref{e2}). However, we expect that such a model selection approach would provide results very similar to those obtained when applying likelihood ratio tests based on mixture likelihoods, already available in the literature.

Regarding the possibility of trimming $\alpha>0$, we now introduce
\begin{equation}\label{xi}
\xi_{P}(\theta,\pi)= \frac{1}
{\sum_{j=1}^k \pi_j p_j(Z_j(\theta,\pi)\cap B(\theta,\pi;P);\theta )}.
\end{equation}
In $\xi_{P}(\theta,\pi)$, the dependence on $P$ is indicated, because $B(\theta,\pi;P)$ does not only depend on $\theta$ and $\pi$, but also on the assumed distribution $P$ since $P[B(\theta,\pi;P)]=1-\alpha$ is required.

Note that $\xi_{P_0}(\theta,\pi)=\eta(\theta,\pi)$ in the untrimmed $\alpha=0$ case because $B(\theta,\pi;P_0) \equiv \mathbb{R}^p$ in that case. In general, for any $\alpha\geq 0$, we have that
$\xi_{P_0}(\theta,\pi)\geq \eta(\theta,\pi),$
and, in particular,
\begin{equation}\label{des_ult}
\xi_{P_0}(\theta^0,\pi^0)= \frac{\eta(\theta^0,\pi^0)}{1-\alpha}.
\end{equation}
Note that the term $\xi_{P_0}(\theta,\pi)$ depends on the probability mass that the normal components with parameters $\theta_j=(\mu_j,\Sigma_j)$ give to the sets $Z_j(\theta,\pi)\cap B(\theta,\pi;P_0)$.

\begin{proposition}\label{pro2} Given $P_0$ with density $f_0$ as in (\ref{e2}), then
\begin{eqnarray}
 &&P_0\bigg[ \sum_{j=1}^{k_0} I_{Z_j(\theta^0,\pi^0) \cap B(\theta_0,\pi_0;P_0)}(\cdot)    \log (\pi_j^0 \phi(\cdot;\theta_j^0)) \bigg] \\ 
 &&- P_0 \bigg[ \sum_{j=1}^k I_{Z_j(\theta,\pi)\cap B(\theta,\pi;P_0)}(\cdot)
    \log (\pi_j \phi(\cdot;\theta_j)) \bigg] \label{pr2_1}\\
    && \geq \log  \frac{\xi_{P_0}(\theta^0,\pi^0)}{\xi_{P_0}(\theta,\pi)}  ,   \label{pr2_2}
\end{eqnarray}
\end{proposition}

\noindent \textit{Proof:} The proof of this result is derived by rewriting the difference in (\ref{pr2_1}) as the sum of terms
$
\mathcal{A}+\mathcal{B}+\mathcal{C}+\mathcal{D},
$
where
\begin{eqnarray*}
 \mathcal{A}&=&P_0\bigg[ \sum_{j=1}^{k_0} I_{Z_j(\theta^0,\pi^0) \cap B(\theta,\pi;P_0)}(\cdot)    \log (\pi_j^0 \phi(\cdot;\theta_j^0)) \bigg] \\
 \mathcal{B}&=&- P_0 \bigg[ \sum_{j=1}^k I_{Z_j(\theta,\pi)\cap B(\theta,\pi;P_0)}(\cdot)
    \log (\pi_j \phi(\cdot;\theta_j)) \bigg]  \\
 \mathcal{C}&=&  P_0 \bigg[ \sum_{j=1}^k I_{Z_j(\theta^0,\pi^0)\cap B(\theta^0,\pi^0;P_0)\cap \overline{B(\theta,\pi;P_0)}}(\cdot)
    \log (\pi_j^0 \phi(\cdot;\theta_j^0)) \bigg]   \\
 \mathcal{D}&=& - P_0 \bigg[ \sum_{j=1}^k I_{Z_j(\theta^0,\pi^0)\cap \overline{B(\theta^0,\pi^0;P_0)}\cap B(\theta,\pi;P_0)}(\cdot)
    \log (\pi_j^0 \phi(\cdot;\theta_j^0)) \bigg].
\end{eqnarray*}
By taking into account the definition of the $B(\theta,\pi;P)$ sets and the fact that
$$
P_0\big[ B(\theta^0,\pi^0;P_0) \cap \overline{B(\theta,\pi;P_0)} \big]=P_0\big[\overline{B(\theta^0,\pi^0;P_0)}\cap B(\theta,\pi;P_0)\big]
$$
(because $P_0[B(\theta^0,\pi^0;P_0)]=P_0[B(\theta,\pi;P_0)]=1-\alpha$), we see that $\mathcal{C}+\mathcal{D}$ is greater or equal than 0. Consequently, we just need to prove that the difference $\mathcal{A}+\mathcal{B}$ is greater or equal than the value in (\ref{pr2_2}).

Let us consider $P_0^*$ being the conditional probability
$$P_0^*(A)=P_0(A|B(\theta,\pi;P_0))=P_0(A\cap B(\theta,\pi;P_0))/(1-\alpha)$$
so that $P_0^*$ has the probability density function
$$f_0^*(x)= \xi_{P_0}(\theta^0,\pi^0,\theta,\pi) \sum_{j=1}^{k_0} I_{Z_j(\theta^0,\pi^0)\cap B(\theta,\pi;P_0)}(x)    \pi_j^0 \phi(x;\theta_j^0)  $$
with
$$
\xi_{P_0}(\theta^0,\pi^0,\theta,\pi)=\frac{1}
{\sum_{j=1}^{k_0} \pi_j^0 p_j(Z_j(\theta^0,\pi^0)\cap B(\theta,\pi;P_0);\theta^0 )}.
$$
Note that 
$$\xi_{P_0}(\theta^0,\pi^0,\theta,\pi)= \frac{\eta(\theta^0,\pi^0)}{1-\alpha}=\xi_{P_0}(\theta^0,\pi^0)$$
as in (\ref{des_ult}) because $P_0[B(\theta,\pi;P_0)]=1-\alpha$.

Finally, take $f^*$ equal to the probability density function
$$f^*(x)=  \xi_{P_0}(\theta,\pi) \sum_{j=1}^{k} I_{Z_j(\theta,\pi)\cap B(\theta,\pi;P_0)}(x)    \pi_j \phi(x;\theta_j),$$
and apply again the Gibbs' inequality to guarantee $P_0^*[\log f_0^*]\geq P_0^*[\log f^*]$. The proof concludes just after rearranging the remaining terms. \hspace*{0.3cm}$\square$ \medskip


As already mentioned,  $\xi_{P_0}(\theta^0,\pi^0)$ and $\xi_{P_0}(\theta,\pi)$ are equal to $\eta(\theta^0,\pi^0)$ and $\eta(\theta,\pi)$, respectively, in the untrimmed $\alpha=0$ case so that Proposition \ref{proposition1} is a particular case of Proposition \ref{pro2}.

If all the $k_0$ components in $P_0$ are well separated and $\alpha>0$, then $\xi_{P_0}(\theta^0,\pi^0)\simeq 1/(1-\alpha)$. In that well-separated case, if we also desire $\xi_{P_0}(\theta,\pi) \simeq 1/(1-\alpha)$ (in order to cancel out approximately the term (\ref{pr2_2}) in   Proposition \ref{pro2}),  it is useful to take $k=k_0$,  $\pi_j \simeq \pi_j^0$ and $\mu_j \simeq \mu_j^0$. However, regarding the scatter matrices, $\Sigma_j \simeq \Sigma_j^0$ is convenient, but with $|\Sigma_j| \leq |\Sigma_j^0|$ (i.e., slightly more concentrated components in terms of $P_0$ are preferred). The following remark is aimed at clarifying this last claim in the simplest case $k=1$, where TCLUST reduces to the well-known MCD. 

\begin{remark}
In the simplest case $k_0=1$, one could expect that an analogue to the inequality (\ref{k_1}) would hold when $P_0$ is the $p$-variate normal with mean $\mu_0$ and covariance matrix $\Sigma_0$. However, we can see that this type of result does not hold. First note that the problem we have to solve to obtain $\mathcal{L}_{\alpha,1}^{\Pi}(P)$, i.e. $k=1$, is exactly the same as the one that underlies the well-known MCD method with an $\alpha$ trimming level when a large $c$ value in the eigenvalue ratio constraint in (\ref{e_1}) is taken (see details in Appendix \ref{app1}). Although the constraint depending on $c$ could be imposed if desired, note that $c$ is no longer needed to avoid degeneracy when $k=1$.

Results in \cite{butler1993asymptotics} together with the material in Appendix \ref{app1} show that
$$
P_0\bigg[\log\phi(\cdot;\mu_0,\widetilde{\Sigma}_0) I_{Z_1}(\cdot)\bigg] \geq P_0 \bigg[ \log (\phi(\cdot;\mu,\Sigma)) I_{Z}(\cdot) \bigg]
$$
for any other $\mu$, $\Sigma$ and $Z$ with $P_0[Z]=1-\alpha$, when $Z_1=B(\theta^0,\pi^0;P_0)=\{x:(x-\mu_0)'\Sigma_0^{-1}(x-\mu_0)\leq \chi^2_{p,\alpha}\}$,  and $\widetilde{\Sigma}_0=\nu_{\alpha}\Sigma_0$ with $\nu_{\alpha}=F_{\chi^2_{p+2}}(\chi^2_{p,\alpha})/(1-\alpha)< 1$. In the previous expressions, $F_{\chi^2_{p}}$ denotes the distribution function of the $\chi^2_p$ distribution, and $\chi^2_{p,\alpha}$ is its $(1-\alpha)$-quantile. The use of the $\nu_{\alpha}^{-1}$ ``correction factor'' is common when applying MCD in practice.

The appearance of the constant  $\nu_{\alpha}<1$ could be understood by noticing that $\xi_{P_0}(\theta^0,1)=1/(1-\alpha)$,  but  $\xi_{P_0}(\tilde{\theta}_0,1) < 1/(1-\alpha)$ for $\theta_0=(\mu_0,\Sigma_0)$, and $\widetilde{\theta}_0=(\mu_0,\widetilde{\Sigma}_0)$, with $|\widetilde{\Sigma}_0|< |\Sigma|$, by making use of Proposition \ref{pro2}. 

\end{remark}

\section{The parametric bootstrap procedure}\label{s3}

The previous section has motivated the investigation of $t^n_{k,\alpha}$ as estimate of $\mathcal{L}_{\alpha,k+1}^{\Pi}(P) - \mathcal{L}_{\alpha,k}^{\Pi}(P)$, which should be close to zero,  in order to make statements about reasonable numbers of clusters. However, sample variability has to be addressed. This makes it difficult to propose automated procedures for choosing $\alpha$ and $k$. In this section, we introduce a new parametric bootstrap approach for deriving more (but not fully) automated choices of parameters for TCLUST as an alternative to the subjective decisions required by visual inspection. Parametric bootstrap was used in a similar manner, but using a different statistic, to determine the number of clusters with the OTRIMLE approach for robust clustering, see \cite{HenCor22}.

Here is the proposed parametric bootstrap procedure. When computing $t_{k,\alpha}^n$ from a sample $\mathcal{X}_n$ of size $n$, we also obtain $\widehat{\pi}_j=\widehat{\pi}_j(\mathcal{X}_n;\alpha,k)$, $\widehat{\theta}_j=(\widehat{\mu_j},\widehat{\Sigma_j})=\widehat{\theta}_j(\mathcal{X}_n;\alpha,k)$ and $\widehat{R_0}=\widehat{R_0}(\mathcal{X}_n;\alpha,k)$, where $\widehat{R_0}$ is the set of trimmed observations. We use the ``hat'' notation to denote the optimal parameters and the optimal partition that achieves the maximum in (\ref{e1}) for the given sample $\mathcal{X}_n$. Taking advantage of this information, we can draw $B$ bootstrap samples
$$\{\mathcal{X}_n^{*b}\}_{b=1}^B \text{ with }\mathcal{X}_n^{*b}=\{x_1^{*b},...,x_n^{*b}\},$$
where the $x_i^{*b}$ values are i.i.d. random realizations from the mixture
$$
\sum_{j=1}^k \widehat{\pi}_j N_p(\widehat{\mu}_j,\widehat{\Sigma}_j) \quad \text{ if }\quad i\notin \widehat{R_0},
$$
and
$$
x_i^{*b}=x_i \quad \text{ if }\quad i\in \widehat{R_0}.
$$
Outliers can be alternatively generated in such a way that they randomly (uniformly) fall in the range of the analysed variables and such that their squared Mahalanobis distances from the centroids $\widehat{\mu}_j$ (computed with the scatter matrices $\widehat{\Sigma}_j$) are larger than a fixed percentile of the chi-squared distribution.

The idea here is to simulate the sampling distribution of $t_{k,\alpha}^n$ in case the data  indeed stem from a Gaussian mixture with $k$ components and a proportion of $\alpha$ of outliers. Without having distribution theory for $t_{k,\alpha}^n$, the parametric bootstrap simulates this sampling distribution based on the Gaussian mixture estimated from the data.

We would want to reject a model defined by choices of $k$ and $\alpha$ in case that $\mathcal{L}(\alpha,k+1;\mathcal{X}_n^{*b})$ is better than $\mathcal{L}(\alpha,k;\mathcal{X}_n^{*b})$ by an amount that cannot be explained by random variation in case the model with $k$ and $\alpha$ is actually true, namely in case that the bootstrap $p$-value
\begin{equation}\label{pkalpha}
p_{k,\alpha}= \frac{\#\{ b:  \mathcal{L}(\alpha,k+1;\mathcal{X}_n^{*b}) - \mathcal{L}(\alpha,k;\mathcal{X}_n^{*b}) > t_{k,\alpha}^n \}}{B}
\end{equation}
is too small. This compares the number of clusters $k$ and $k+1$ for given $\alpha$. The interpretation of rejection is that $k$ is not a good and rather too low value for the number of clusters for given $\alpha$ as increasing $k$ clearly improves $\mathcal{L}(\alpha,k;\mathcal{X}_n^{*b})$; conversely, non-rejection would make $k$ a candidate number of clusters, as potential benefits of increasing $k$ by 1 can be explained by random variation. Experience suggests that in case $k$ is chosen too low, indeed a small $p$-value will usually be observed, even if $k+1$ is still too small, i.e., in case that two inappropriate solutions are compared, increasing $k$ normally still improves matters significantly. Although the procedure does not directly compare different values of $\alpha$, it provides some information on the choice of $\alpha$. In particular, if the same $k$ looks like a suitable number of clusters for a larger and a smaller $\alpha$, it would normally be preferred to put more observations into clusters, i.e., to choose the smaller $\alpha$.

Therefore, the sequential procedure described in pseudo-code in Algorithm \ref{alg1} can be used to choose sensible values for $k$ and $\alpha$. We can declare $p_{k,\alpha}$ as small if $p_{k,\alpha}< \texttt{crit}$ for a small fixed $\texttt{crit}$ value (for instance, \texttt{crit}$=0.1$).

The algorithm works as follows:

\hrulefill
\begin{algorithm}[H]
 \KwData{The sample $\mathcal{X}_n$, the maximum trimming level $\alpha_{\max}$,  the maximum number of clusters $k_{\max}$, the length of the grid $L$, the value \texttt{crit} (e.g. \texttt{crit}=0.1) and a constant $c$ constraining the ratio between eigenvalues of the scatter matrices through $\Theta_c$}
 \KwResult{A list $sensible(1)$, $sensible(2)$,... with sensible pairs $(k,\alpha)$}
 Initialize: $ p_{k,\alpha} \leftarrow  \texttt{Na}$; $\texttt{sol} \leftarrow 0$;
 $k \leftarrow 1$;
 $\alpha \leftarrow 0$;
 $k_{\text{best}} \leftarrow k_{\max}+1$\;
 \While{$\alpha \leq \alpha_{\max}$}{
   \While{$k < k_{\text{best}}$}{
	 Compute $p_{k,\alpha}$ (parametric bootstrap in (\ref{pkalpha}))\;
     \eIf{$p_{k,\alpha} > \emph{\texttt{crit}}$}{
      $\texttt{sol} \leftarrow \texttt{sol}+1$\;
      $sensible(\texttt{sol}) \leftarrow (k,\alpha)$\;
      $k_{\text{best}} \leftarrow k$\;
      }{
      $k \leftarrow k+1$\;
      }
   }
   $\alpha \leftarrow \alpha + \frac{\alpha_{\max}}{L} $\;
   $k \leftarrow 1$\;
 }
 \caption{Selecting $k$ and $\alpha$ }\label{alg1}
\end{algorithm}

%

Even though $sensible(1)$, $sensible(2)$,\ldots are provided in a sequential order, it does not mean that $sensible(1)$ is necessarily the best solution and so on. All the solutions returned can be considered as legitimate, and the user will need to make a final decision if a single clustering is required. Several graphical and numerical tools are proposed in the literature for assessing clusterings (see, e.g., \cite{hennig2005method}), and knowledge regarding the background of the data set at hand or the purposes of clustering can also be involved. We will consider $sensible(1)$ as the preferred solution in the simulation study in Section \ref{s5}. This choice can be justified by noting that smaller sized clusters (smaller than an $\alpha$ portion of the data) can be completely trimmed with large enough $\alpha$. This choice, however, is only a convention, mostly for comparative purposes with other procedures, and it is well conceivable in certain situations that another solution is preferable for the user. Our procedure provides a list of sensible solutions from which the user can pick an optimal one for the given purpose.

There is a possibility for further research and (perhaps) improvements of the proposed approach. Note that only changes on consecutive values of $k$ (changing $k$ to $k+1$) for fixed $\alpha$ are analysed. It could be studied whether considering bootstrap testing to compare directly $k$ against $k'$ could improve the performance. This has not been the case in all of the initial trials we have conducted. Algorithm \ref{alg1} is designed to keep the computational cost as low as possible. It can be modified to return all the $p_{k,\alpha}$ values for all combinations of $k$ and $\alpha$ when $k=1,...,k_{\max}$ and $\alpha=0,...,\alpha_{\max}$. All these values of $p_{k,\alpha}$ could be visualized, as their magnitudes may provide additional valuable information. However, this aspect has not been considered yet. 

\section{Illustrative examples}\label{s4}

We go back to the data set shown in Figure \ref{f1}. This data set consists of $n=1000$ observations generated from four bivariate normally distributed components with location vectors $(3,23)$, $(23,3)$, $(14,14)$, and $(30,30)$, with 240, 230, 200, and 230 observations, respectively. Two additional components were generated, with 20 uniformly distributed observations in the rectangles $[10,15]\times [35,40]$, and 80 uniformly distributed observations in the rectangle $[30,60]\times [25,45]$. 

We apply the proposed methodology with $c=50$, $\alpha_{\max}=0.2$, and $k_{\max}=7$. Table \ref{table:1} shows the values of $p_{k,\alpha}$ when applying Algorithm \ref{alg1} with $B=100$ and $\texttt{crit}=0.1$.

\begin{table}
\centering
\begin{tabular}{rrrrrrrrrr}
 &  &  &  &  & $\alpha$  &  &  &  &  \\
\cline{2-10}
$k$       &  0.000  &  0.025 &  0.050 &  0.075  &  0.100 &  0.125 &  0.150  &  0.175 &  0.200\\\hline
1 &	0.00		 &  0.00	 	     & 	0.00		 & 	0.00		 &  0.00		     & 	0.00		 & 	0.00		 & 	0.00		 & 	0.00	 \\
2 & 	0.00			 &  0.00		     & 	0.00		 & 	0.00		 &  0.00		     & 	0.00		 & 	0.00		 & 	0.00		 & 	0.00	 \\
3 &	0.00			 &  0.00		     & 	0.00		 & 	0.00		 &  0.00		     & 	0.00		 & 	0.00		 & 	0.00		 & 	0.00	 \\
4 &	0.00			 &  0.00		     & 	0.00		 & 	0.00		 &  \textbf{0.73}	     &   \texttt{Na} 	     &    \texttt{Na} 	    &    \texttt{Na} 		 &    \texttt{Na} 	 \\
5 &	0.00			 &  0.00		     & 	0.00		 & 	\textbf{1.00}		 &   \texttt{Na} 	     &   \texttt{Na} 		 &    \texttt{Na} 		 &    \texttt{Na} 		 &   \texttt{Na} 	 \\
6 &	\textbf{0.21}	     &  \texttt{Na} 		 &  \texttt{Na} 	 &  \texttt{Na}  &  \texttt{Na}     & 	  \texttt{Na} 	 &   \texttt{Na} 	 & 	  \texttt{Na} 	 & 	  \texttt{Na} \\
7 &	  \texttt{Na} 	     &   \texttt{Na} 	 &   \texttt{Na}  &   \texttt{Na} 	 & 	  \texttt{Na} 		 &   \texttt{Na}   &   \texttt{Na} 	 &   \texttt{Na} 	 &   \texttt{Na} \\
\hline
\end{tabular}


\caption{Values of $p_{k,\alpha}$ for the data set in Figure \ref{f1}(a). The values in bold-face correspond to the three solutions exhibited in Figure \ref{f2} (b), (c) and (d).}\label{table:1}
\end{table}

\begin{figure}
\centering
\hspace*{0.5cm}{\scriptsize \textsf{(a)}} \hspace*{5.75cm}{\scriptsize \textsf{(b)}}

\includegraphics[clip,width=6.25cm, height=5.25cm]{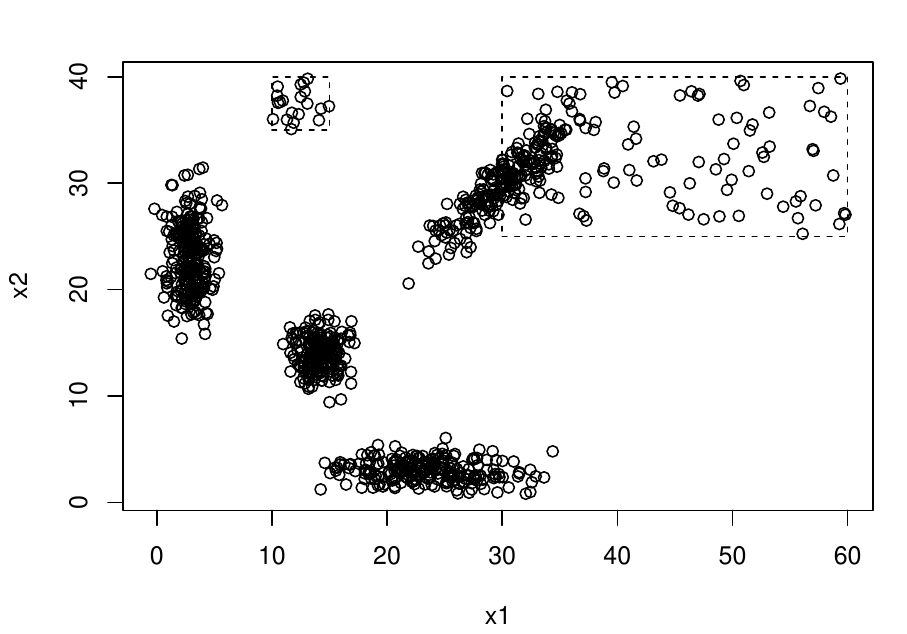} \includegraphics[clip,width=6.25cm, height=5.25cm]{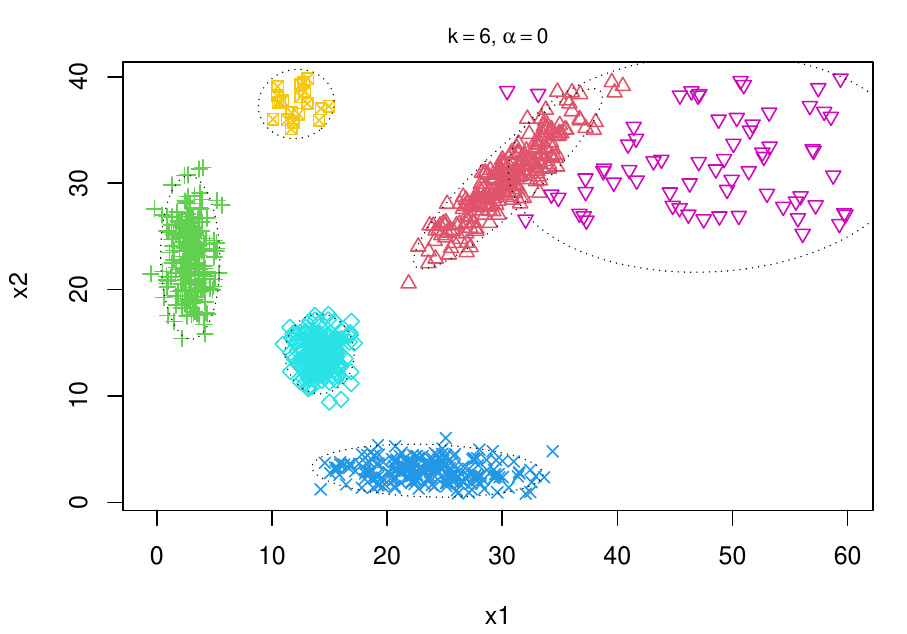}

\hspace*{0.5cm}{\scriptsize \textsf{(c)}} \hspace*{5.75cm}{\scriptsize \textsf{(d)}}

\includegraphics[clip,width=6.25cm, height=5.25cm]{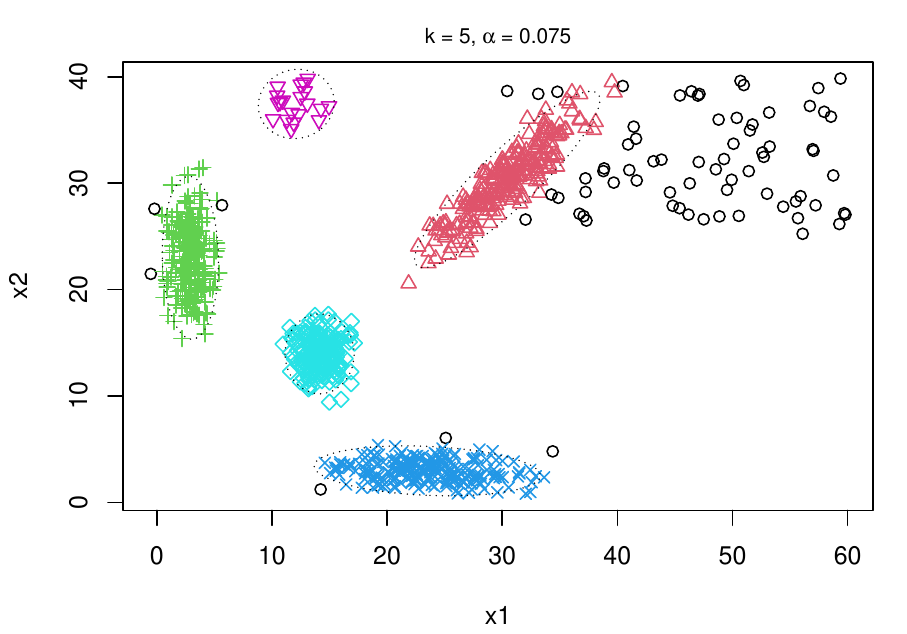} \includegraphics[clip,width=6.25cm, height=5.25cm]{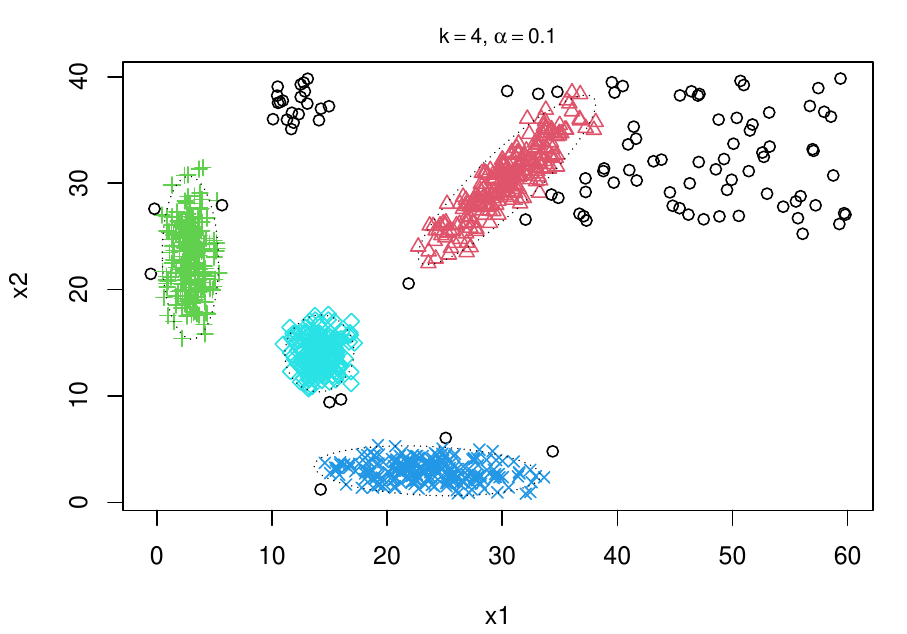}

\caption{The data set in Figure \ref{f1}(a) with the groups of points the cluster nature of which is more ambiguous highlighted in (a). The three sensible choices returned for $k$ and $\alpha$ are shown in (b), (c), and (d) when $k_{\max}=7$ and $\alpha_{\max}=0.2$, which result from Table \ref{table:1}.}\label{f2}
\end{figure}

The $p_{k,\alpha}$ values in Table \ref{table:1} suggest three sensible choices of $k$ and $\alpha$. The corresponding clusterings are shown in Figure \ref{f2}. These corresponding choices are $k=6$ and $\alpha=0$ in (a), $k=5$ and $\alpha=0.05$ in (b) and $k=4$ and $\alpha=0.1$ in (c). The two less dense and not normally distributed components are the observations trimmed by the solutions (b) and (c).

\begin{figure}
\centering
\hspace*{0.5cm}{\scriptsize \textsf{(a)}} \hspace*{5.75cm}{\scriptsize \textsf{(b)}}

\includegraphics[clip,width=6.25cm, height=5.25cm]{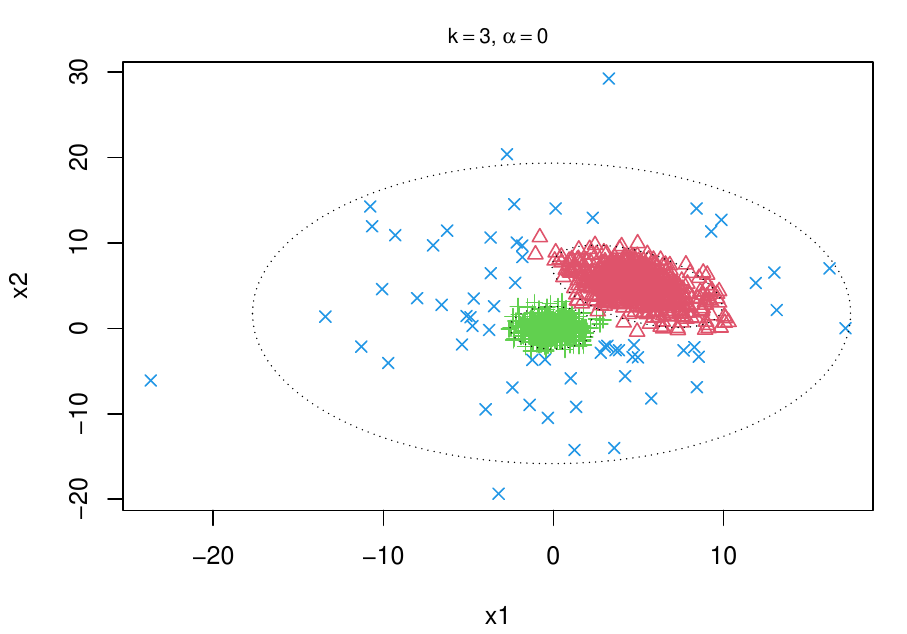} \includegraphics[clip,width=6.25cm, height=5.25cm]{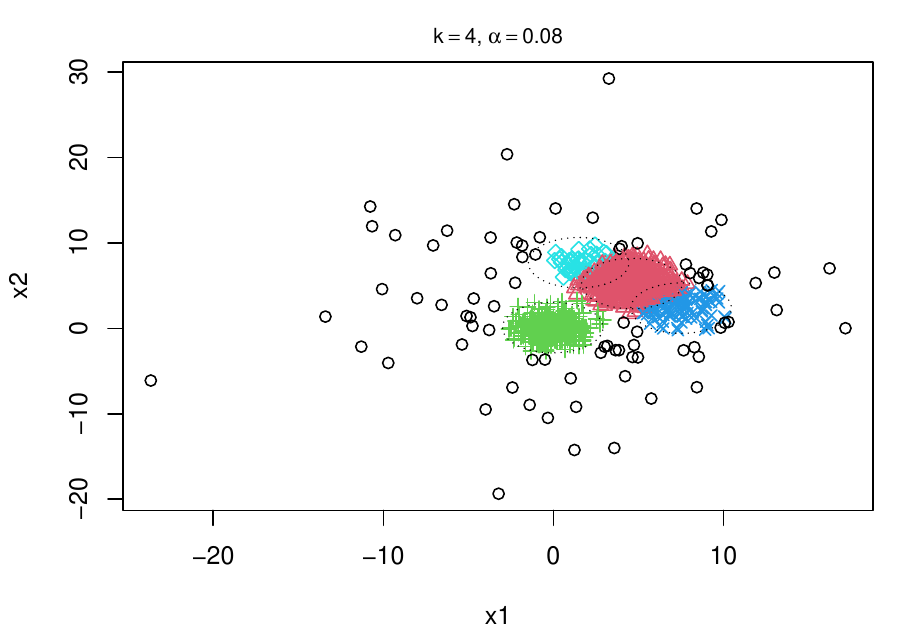}

\caption{Artificial data with one of the sensible solutions found using $c=50$  in (a) and one of the sensible solutions using $c=1$ in (b).}\label{f3}
\end{figure}

We now show an example that illustrates the importance of the determination of the eigenvalue ratio constraint $c$. The data set in Figure \ref{f3} is made of three normally distributed components. One of them has far larger within-cluster variation than the others. Figure \ref{f3}(a) shows one of the selected sensible solutions using $c=50$, and (b) shows one of the selected solutions using $c=1$. Using a large eigenvalue ratio constraint such as $c=50$ allows the detection of the most scattered component. This solution is no longer found as a sensible solution with $c=1$. On the other hand, the choice of $c=1$ makes the solution shown in Figure \ref{f3}(b) sensible, but this solution is not considered as sensible with $c=50$. Recall that $c=1$ forces the detected clusters to be spherical and equally scattered. The solution (a) corresponds better to the way how the data were generated. But solution (b) could be preferred in applications in which clusters are required to have low within-cluster (Euclidean) distances, as is the case for example in social stratification, see \cite{HenL03}. The user needs to specify the allowed differences in variabilities among clusters, which has an impact on the kind of cluster that can be found by the method. Subject matter knowledge can be involved here, and full automation is not necessarily desirable even if it could be achieved.

\section{Simulation study}\label{s5}

We have emphasized that the sensible solution detected first should not necessarily always be the preferred one, but, for the sake of comparison, we consider this as an ``automated'' choice in the simulation study. Recall that this first solution is associated with the smallest trimming size among the sensible solutions. The proposed bootstrap method is referred to here as \textsf{Boot-tclust}.

\begin{figure}
\centering

\hspace*{0.5cm}{\small \textsf{$K=2$ and $p=2$}}
\includegraphics[clip,width=12cm, height=9cm]{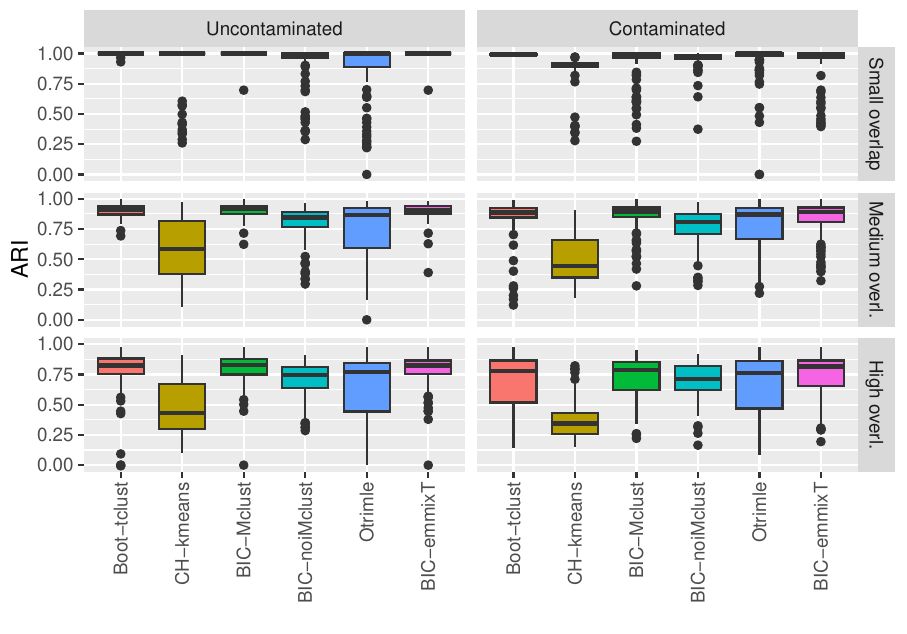}

\hspace*{0.5cm}{\small \textsf{$K=2$ and $p=6$}}
\includegraphics[clip,width=12cm, height=9cm]{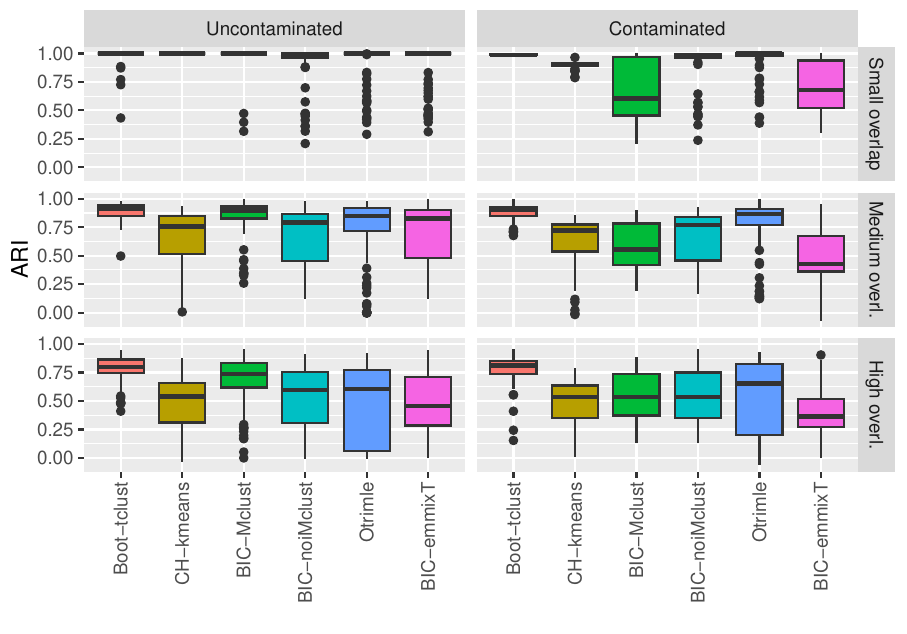}

\caption{Comparison of the ARI values achieved by various clustering methods combined with methods to estimate the number of clusters (the proposed method is denoted LRT) for 100 simulated data sets with real number of clusters $K=2$ with dimensions $p=2$ (top) and $p=6$ (bottom).}\label{f4a}
\end{figure}

\begin{figure}
\centering

\hspace*{0.5cm}{\small \textsf{$K=4$ and $p=2$}}
\includegraphics[clip,width=12cm, height=9cm]{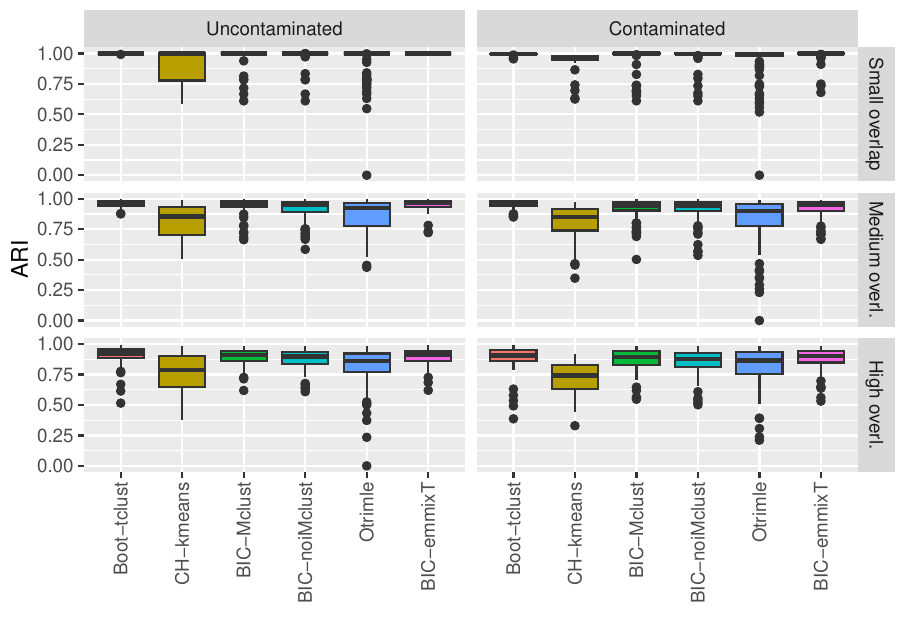}

\hspace*{0.5cm}{\small \textsf{$K=4$ and $p=6$}}
\includegraphics[clip,width=12cm, height=9cm]{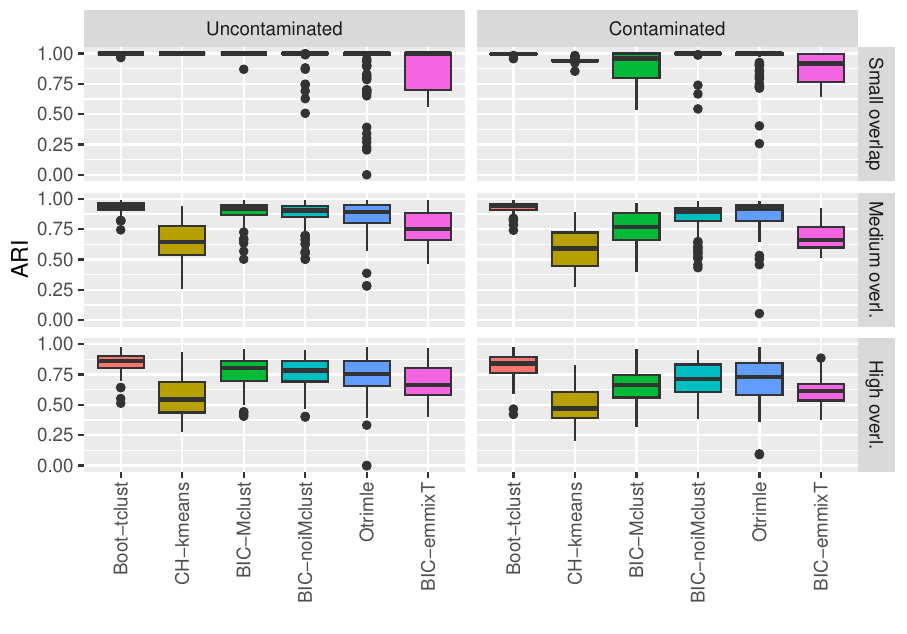}

\caption{Comparison of the ARI values achieved by various clustering methods combined with methods to estimate the number of clusters (the proposed method is denoted \textsf{Boot-tclust}) for 100 simulated data sets with real number of clusters $K=2$ with dimensions $p=4$ (top) and $p=6$ (bottom).}\label{f4b}
\end{figure}

We compare its clustering performance with respect to other automated choices of the clustering. We compare it with the Calinski-Harabasz index \citep{calinski1974dendrite} combined with the classical $k$-means (denoted as \textsf{CH-kmeans} in Figure \ref{f4a} and \ref{f4b}), the BIC criterion implemented in the \texttt{mclust} package \citep{ScrM16}  (denoted as \textsf{BIC-Mclust}) and also its version including a noise component (denoted as \textsf{BIC-noiMclust}; \cite{FraRaf98}), the OTRIMLE approach \citep{coretto2016robust,otrimle} for robust-clustering with an improper uniform distribution (denoted as \textsf{Otrimle}), using the parametric bootstrap scheme introduced in \cite{HenCor22}, and the use of mixtures of $t$-distributions (denoted as \textsf{BIC-emmixT}) implemented in the \texttt{EMMIX} package.

We generate 100 samples in dimensions $p=2$ and $p=6$ of $K$ normally distributed components with $K=2$ and $K=6$ with different degrees of overlap and such that the constraint for the corresponding scatter matrices (\ref{e_1}) is fulfilled with $c=12$. Each of the $K$ components includes 100 observations generated by applying the extension of the \texttt{MixSim} method of \cite{22c} given in  \cite{24b}. This extension allows to impose the desired  eigenvalue ratio constraint and also to impose prespecified overlap rates using an \texttt{overlap} parameter. Given two clusters $j$ and $l$ obtained from normal densities $\phi(\cdot;\mu_j, \Sigma_j )$ and $\phi(\cdot;\mu_l, \Sigma_l)$, with probabilities of occurrence $\pi_j$ and $\pi_l$, the overlap between groups $j$ and $l$ is defined as the sum of the two misclassification probabilities $w_{jl} = w_{j|l} + w_{l| j}$ where $w_{j|l} = P[\pi_l\phi(X;\mu_l, \Sigma_l) < \pi_j\phi(X;\mu_j, \Sigma_j)]$.
Three levels of average overlap are considered by setting the \texttt{overlap} parameter in \cite{24b} equal to 0 (\textsf{No overlap}), 0.05 (\textsf{Medium overl.}), and 0.1 (\textsf{High overl.}). This simulation scheme corresponds to the \textsf{Uncontaminated} case. For the data sets denoted as \textsf{Contaminated} we add another 10 observations (when $K=2$), another 20 observations (when $K=4$), respectively, that are uniformly distributed in the range defined by the non-contaminated data.

Figure \ref{f4a} and \ref{f4b} show boxplots of the ARI index values \citep{HubA85} for the partitions obtained after applying all the compared procedures to the same 100 simulated data sets, compared with the true data generation scheme. The ARI does not allow for noise or trimmed observations. In order to compute these ARI values, we consider observations generated from the uniform as an additional cluster in the \textsf{Contaminated} case regarding the true data generation process. Consequently, the observations labeled as noise or trimmed (for methods allowing that possibility) are also evaluated as an additional cluster.

We can see that all methods, apart from \textsf{CH-kmeans}, which is not well suited to deal with non-spherical clusters as generated here, perform quite well in the \textsf{No overlap case}. However, the proposed bootstrap method \textsf{Boot-tclust} outperforms the competitors in the overlapped cases.

\begin{figure}
\centering

\hspace*{-2cm}{\small \textsf{$p=2$}}

\includegraphics[clip,width=12cm, height=9cm]{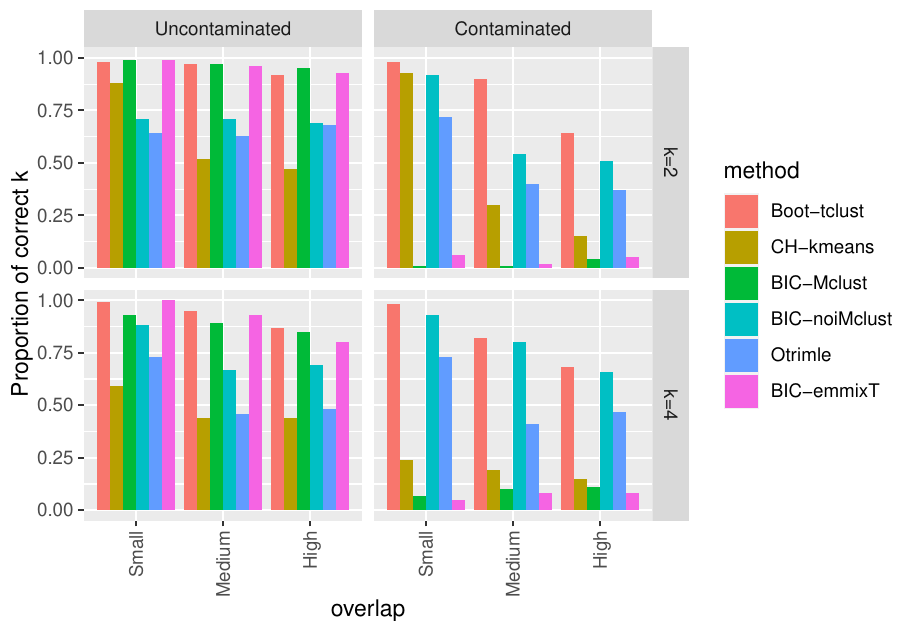}

\hspace*{-2cm}{\small \textsf{$p=6$}}

\includegraphics[clip,width=12cm, height=9cm]{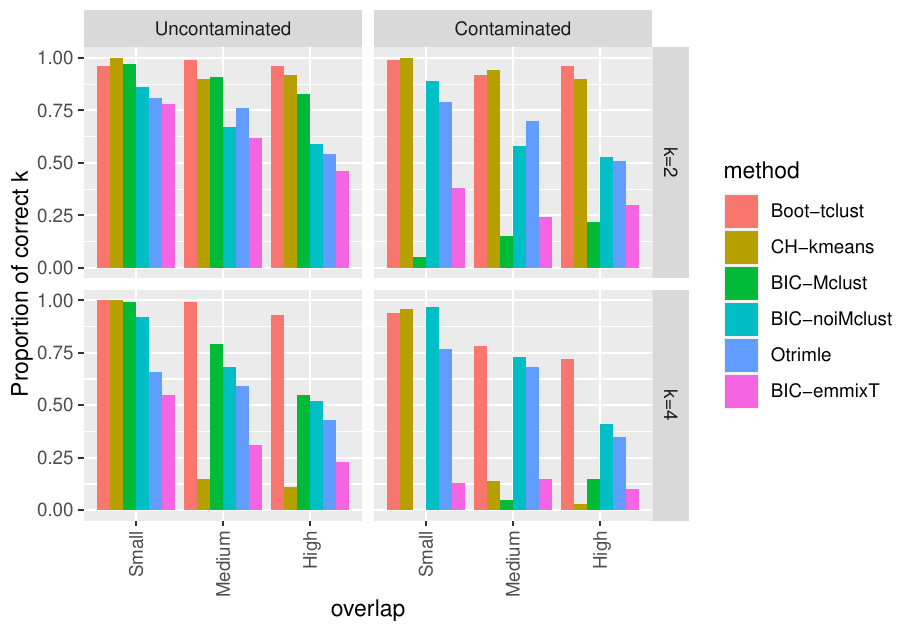}

\caption{Proportion of times that the true number of clusters is detected in dimension $p=2$ (top) and dimension $p=6$ (bottom).}\label{f40}
\end{figure}

For complementing the simulation study, Figure \ref{f40} shows the proportion of times that the true number of clusters is detected in dimension $p=2$ and dimension $p=6$, again when choosing the sensible solution detected first. We compare the results of the proposed methodology \textsf{Boot-tclust} with respect to the other approaches already mentioned. Although these other procedures perform reasonably well in the \textsf{Uncontaminated} case with \textsf{No overlap}, the performance of \textsf{Boot-tclust} is consistently better in the \textsf{Contaminated} where only \textsf{BIC-noiMclust} and \textsf{Otrimle} can provide sensible results, but no accurate as \textsf{Boot-tclust} when overlap increases.

\section{Real data examples}\label{s6}

\subsection{Old Faithful Geyser data}\label{s6_1}

\begin{figure}
\centering

\centering
\hspace*{0.5cm}{\scriptsize \textsf{(a)}} \hspace*{5.7cm}{\scriptsize \textsf{(b)}}

\includegraphics[clip,width=6.25cm, height=5.25cm]{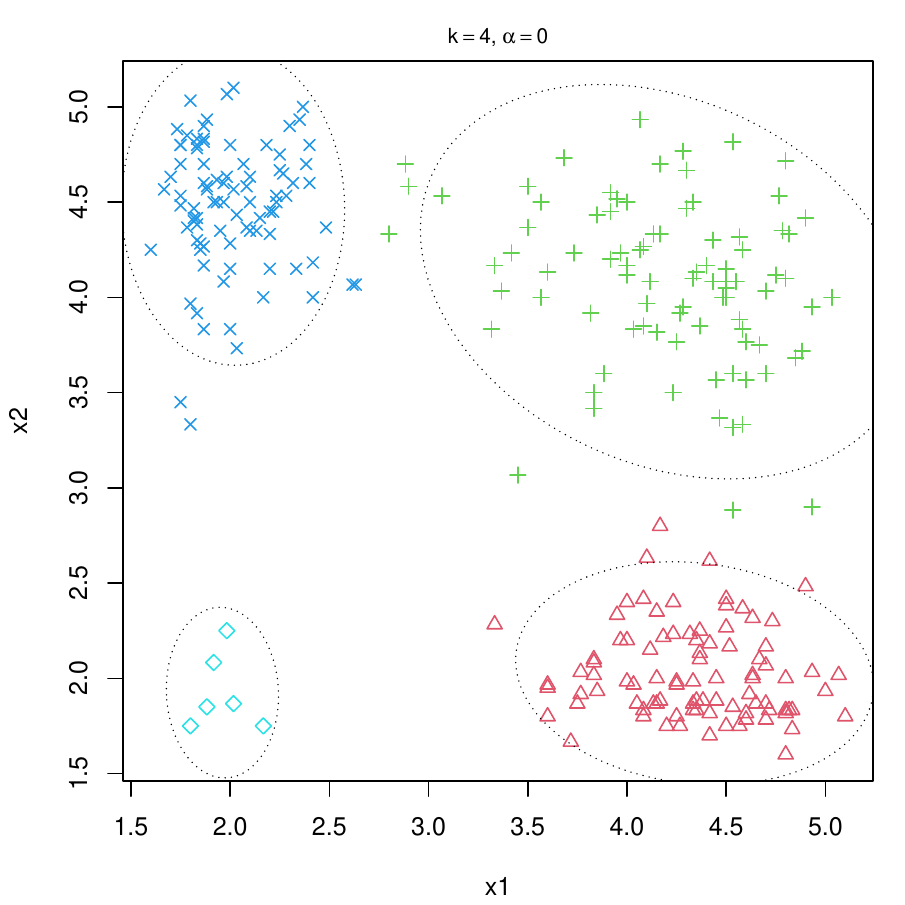} \includegraphics[clip,width=6.25cm, height=5.25cm]{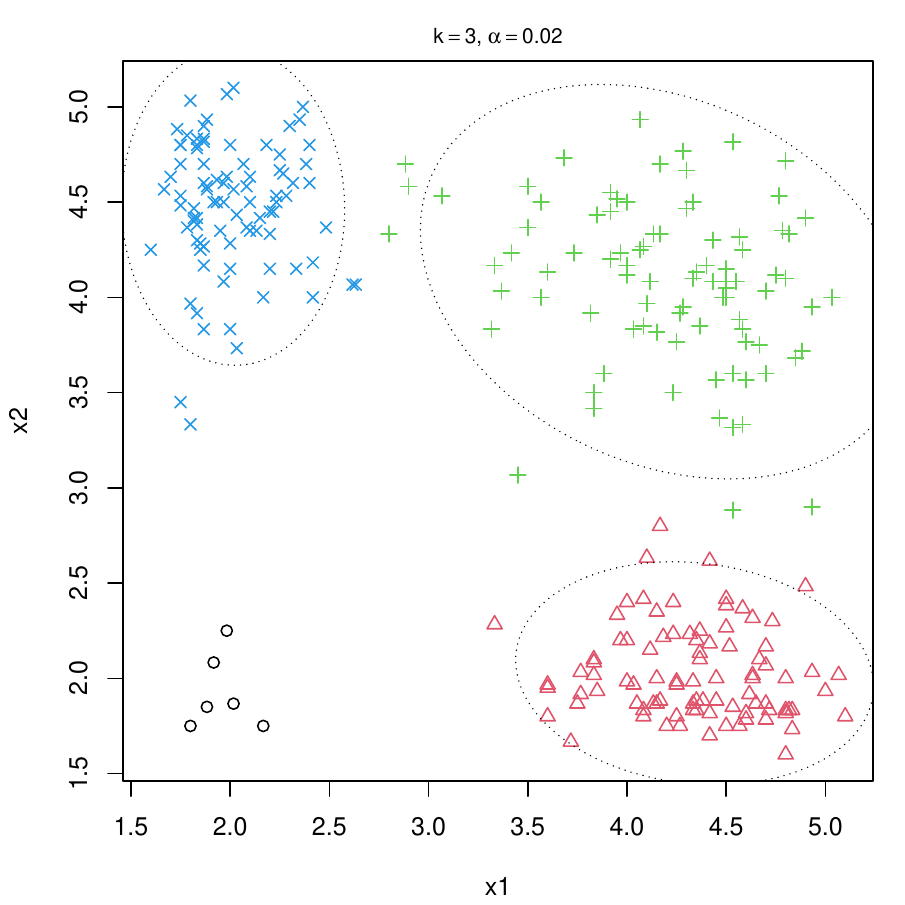}

\caption{The two sensible solutions obtained for the Old Faithful Geyser data example.}\label{f5}
\end{figure}

As real data example, we first consider a simple bivariate data set. The data set contains the eruption length and the length of the previous eruption in minutes for 271 eruptions of the Old Faithful Geyser in Yellowstone National Park. This data set accompanies both the \texttt{tclust} package and the \texttt{FSDA} toolbox as the \texttt{geyser2} data set (obtained from \cite{Hrdle1991}, p. 34).
The proposed methodology with $k_{\max}=7$, $\alpha\in\{0,0.02,0.04,...,0.18,0.2\}$, and $c=50$ delivers the two sensible solutions shown in Figure \ref{f5}(a) and (b). It could be debatable if the small group of ``short-followed-by-short'' eruptions (the six observations in the left-down corners of these figures) should be considered as a proper cluster or rather as a small fraction of anomalous eruptions. Ultimately the user has to decide this unless they are happy with two solutions. In any case, other no so sensible combinations of $k$ and $\alpha$ are ruled out by the method.

\subsection{COVID data}\label{s6_2}

\begin{figure}
\centering

\includegraphics[clip,width=10cm, height=8cm]{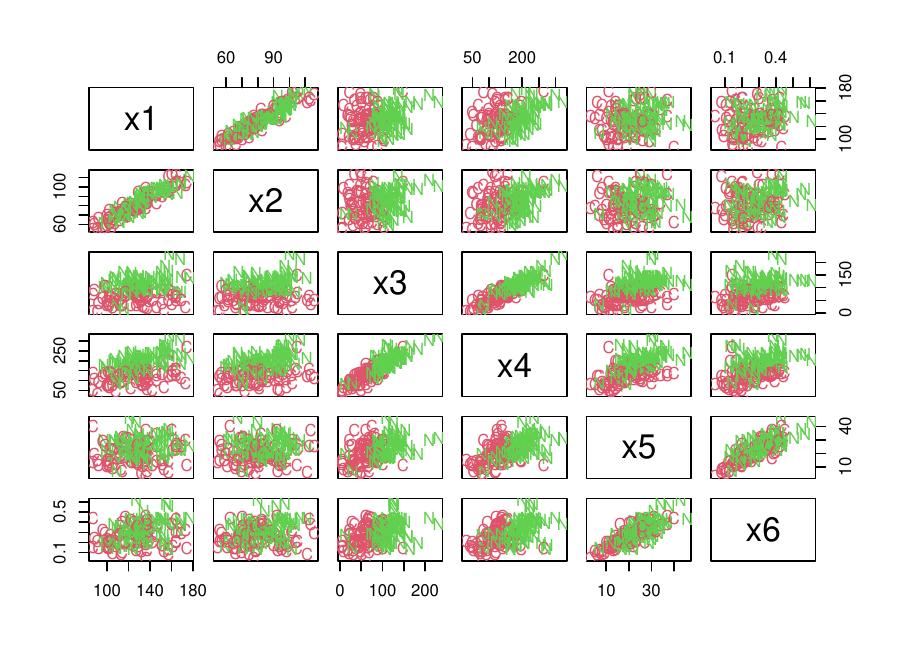}

\caption{Pairs plot for the COVID data set. Symbols distinguish between patients who claim to have had COVID (\texttt{C} symbol) and those who have not (\texttt{N} symbol).}\label{covid_a}
\end{figure}

The second example regards the analysis of a data set on the SARS-CoV-2 symptoms kindly provided to us by the ASL3 Genovese Hospital. Measurements on 6 variables were used, namely $x_1:$ ``heart rate (the number of beats the heart per minute)'',
$x_2:$ ``Oxygen Uptake Efficiency Slope (index of functional reserve derived from the logarithmic relation between oxygen uptake and minute ventilation during incremental exercise)'', $x_3:$ ``watts (reached by the patient during the stress test on a cycle ergometer on a stationary bike at the aerobic threshold, that is, when the patient 'begins to struggle')'', $x_4:$ ``watts peak (watts reached at maximum effort (during exercise test on exercise bike)'', $x_5:$ ``value of the maximum repetition (maximum force of muscle contraction of the quadriceps femoris of the dominant limb expressed in kg)'', and $x_6:$ ``previous variable corrected on the subject (in relation to the patient's weight)'' on 79 patients that claim to have gone through a COVID infection, and another 77 who claim not to have gone through it.
Note that when we write ``claim to have gone through'', this means that in some cases the classification is based on the declaration of the patient without any supporting document. Moreover the date on which the patients declared to have had COVID is not available. 
Finally, it could be that some patients did have COVID but were unaware of it.
This dataset is a mixture between discrimination and classification because we know a priori that there is surely more than one group, but it is not clear whether the dichotomous classification by the doctors (which is partially based on the patient declarations) is too rough. Moreover, the role of the different variables in the separation among the groups is unclear.
Data have been collected by the Post-COVID Outpatient Rehabilitation Center ASL3 Liguria Region Health System and  approved by the Ethics Committee of the region of Liguria (Italy). Figure \ref{covid_a} shows a pairs plot 
by using the symbols \texttt{C} (COVID reported by the patient) and \texttt{N} (COVID not reported by the patient). 


\begin{figure}

\centering
\hspace*{0.5cm}{\scriptsize \textsf{(a)}} \hspace*{5.7cm}{\scriptsize \textsf{(b)}}
\includegraphics[clip,width=6.25cm, height=6.75cm]{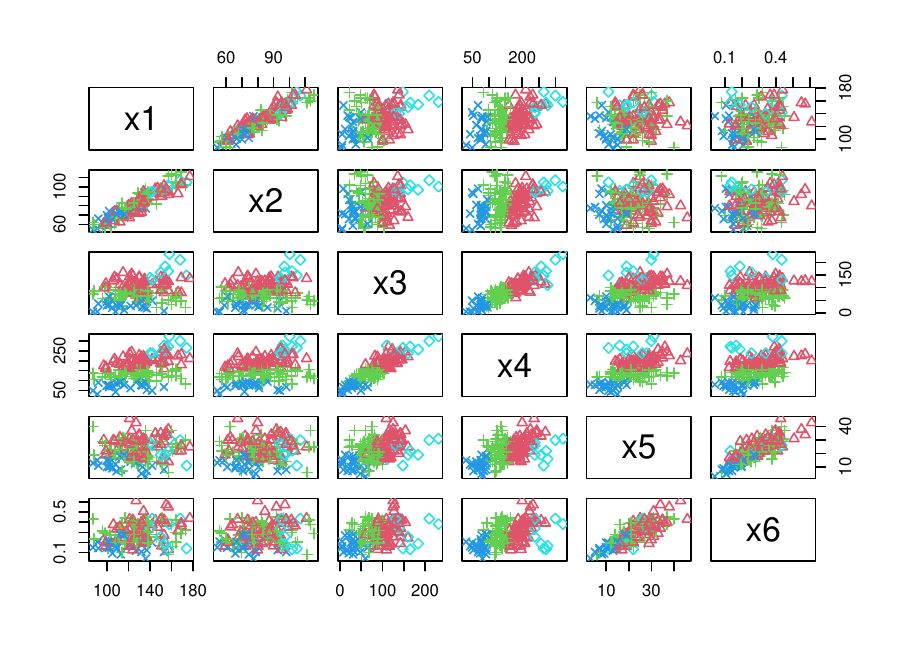} \includegraphics[clip,width=6.25cm, height=6.75cm]{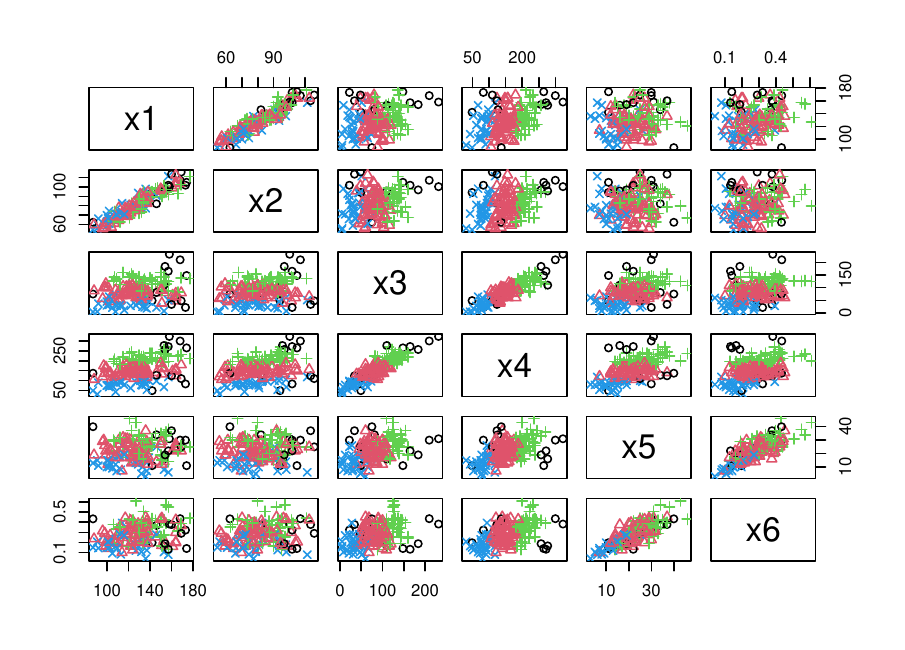}

\caption{Pairs plot showing the two sensible solutions obtained for the COVID data set.}\label{covid_b_c}
\end{figure}

When only using the six numerical variables (and not the patients' claims), we find the two sensible solutions shown in Figure \ref{covid_b_c}(a) and (b). These two solutions were selected by the proposed parametric bootstrap procedure with $k_{\max}=7$, $\alpha_{\max}=0.2$, $c=100$, and $\texttt{crit}=0.1$. The chosen parameters are $k=4, \alpha=0$, and $k=3, \alpha=0.075$, respectively. They mainly differ in whether a certain fraction of observations is trimmed as outlying or estimated as a proper fourth cluster. These observations seem to depart from the pattern followed by the others ones.
Just focusing on the clusters corresponding to the blue ``$\times$'' symbols and the green ``$+$'' symbols in Figure \ref{covid_b_c}(b), we could see that the first of these two clusters includes 44 patients of which 42 have reported COVID, and the second one includes 31 patients of which 29 have not reported COVID. Therefore, these two clusters (blue ``$\times$'' and the green ``$+$'' symbols) identify groups of patients. These groups correspond well to the patients' claims.  The intermediate cluster with red ``$\triangle$'' symbols in Figure \ref{covid_b_c}(b) includes cases that are compatible with both the ``N'' and the ``C'' class. An advantage of the solution in Figure \ref{covid_b_c}(b) is that trimming allows to find heterogeneous outliers, whereas only a more homogeneous part of these is in the ``$\diamond$''-cluster in Figure \ref{covid_b_c}(a). Although clustering treats the data set as heterogeneous and will therefore in principle allow for outliers to form their own clusters, it is hard to estimate the parameters for very small clusters, and trimming such observations and treating them as outliers will often work better for data subsets that are small, heterogeneous, and deviate clearly from the larger clusters.

\section{Conclusions and further directions}\label{s7}

We have provided some theoretical justification for the application of the \texttt{ctlcurves} approach introduced in \cite{GarG11} for choosing a number of clusters and trimming rate. A novel parametric bootstrap approach has also been introduced, aimed at an automatic and principled assessment of these \texttt{ctlcurves}. The parametric bootstrap procedure equips the user with a small set of sensible solutions. The user then still needs to decide whether and which of these solutions works well for the purpose of clustering in the situation at hand. This decision can be made by taking into account background information regarding the data and the aim of clustering, visualization, and cluster validation using available tools.

The returned set of sensible solutions is dependent on the choice of the covariance eigenvalue ratio constraint $c$. The parameter $c$ implies a decision regarding what shapes of clusters are admitted. For instance, a value of $c$ close to 1 will detect clusters close to sphericity and with similar scatters. On the other hand, the constant $c$ should be chosen not so small (but finite) if the user does not require very constrained types of clusters. Anyway, our experience is that the determination of the sensible solutions is often very stable and not very dependent on the specific choice of $c$ as long as it is not very close to 1 or very large (see the discussion in \cite{GarG11}). For a more general approach regarding the constraints for the scatter matrices see \cite{garcia2022constrained}.

A further ingredient to take into account when choosing $k$ and $\alpha$ is the assessment of the stability of contiguous optimal partitions when moving $k$ and $\alpha$ in a controlled way. This could be a topic for future research. Monitoring stability was already applied in \cite{CerG18} and, more recently, in \cite{torti2021semiautomatic} and \cite{cappozzo2023graphical}. Another interesting idea could be the development of graphical summaries for taking advantage of all the bootstrap $p$-values.

There is also room for improvement regarding computational aspects since the introduced bootstrap approach is a computationally demanding procedure due to its dependence on TCLUST. Trying to reduce computing times, for instance, it could make sense to take into consideration the already available solutions, obtained for other $k$ and $\alpha$ values, to be tried out as tentative initial solutions together with the random initializations that are a standard part of TCLUST.

The proposed methodology can be implemented after installing the \texttt{FSDA} MATLAB toolbox, which can be obtained either from the MathWorks file exchange or from GitHub at \url{https://github.com/UniprJRC/FSDA/}. The main function which computes the \texttt{ctlcurves} together with the parametric bootstrap $p$-values is \textsf{ctlcurves.m}. The two datasets used in Section \ref{s6} are also available from this MATLAB toolbox (\textsf{geyser2} and \textsf{covid} datasets).

\appendix
\section{Equivalence between TCLUST and MCD in the case $k=1$}\label{app1}
In the case $k=1$, removing the eigenvalue ratio constraint (by taking a very large $c$ value in  (\ref{e1})), the maximization done to obtain $\mathcal{L}_{\alpha,1}^{\Pi}(P)$ reduces to the maximization of
\begin{equation}\label{equiv}
 P \bigg[ \log (\phi(\cdot;\mu,\Sigma)) I_{Z}(\cdot) \bigg],
 \end{equation}
where $P[Z]=1-\alpha$.

If the set $Z$ were fixed, the straightforward adaptation of the classical maximum likelihood theory under multivariate normality tells us that the optimal $\mu$ and the optimal $\Sigma$, for that fixed $Z$, should be $\mu=\mu(Z)$ and $\Sigma=\Sigma(Z)$ where
$$
\mu(Z)=\frac{1}{1-\alpha}\int_{Z}xdP(x)
$$
and
$$
\Sigma(Z)=\frac{1}{1-\alpha}\int_{Z}(x-\mu(Z))(x-\mu(Z))'dP(x).
$$
By taking into account that
$$
-2\log \phi(x;\mu,\Sigma)=p\log(2\pi)+\log|\Sigma|+(x-\mu)'\Sigma^{-1}(x-\mu),
$$
the maximization of (\ref{equiv}) on $\mu$, $\Sigma$ and $Z$ reduces to the minimization on $Z$ of
$$
(1-\alpha)p\log(2\pi)+(1-\alpha)\log|\Sigma(Z)|+ \int_{Z}(x-\mu(Z))' \Sigma(Z)^{-1}(x-\mu(Z))dP(x).
$$
This last minimization finally boils down to the minimization of $|\Sigma(Z)|$ (i.e., the MCD population objective function) since
\begin{eqnarray*}
&&\int_{Z}(x-\mu(Z))' \Sigma(Z)^{-1}(x-\mu(Z))dP(x)= \\
&&=\int_{Z}\text{trace}((x-\mu(Z))' \Sigma(Z)^{-1}(x-\mu(Z)))dP(x) =(1-\alpha)p,
\end{eqnarray*}
where the last inequality follows from the cyclic property of the ``trace'' function.

\bibliographystyle{abbrvnat}
\bibliography{refs6}

\end{document}